\documentclass[aps,prb,twocolumn,superscriptaddress,showpacs,floatfix,longbibliography,10pt]{revtex4-2}
\usepackage[english]{babel}
\usepackage{fullpage}
\usepackage{amssymb}
\usepackage{graphicx,color}
\usepackage{bbm}
\usepackage{multirow}
\usepackage{bm}
\usepackage{mathrsfs}
\usepackage{commath}
\usepackage{stmaryrd}
\usepackage{color}
\usepackage{subfigure}
\usepackage{comment}
\usepackage[utf8]{inputenc}
\usepackage[normalem]{ulem}
\usepackage{amsmath}
\usepackage{graphicx}
\usepackage{float}
\usepackage{verbatim}
\usepackage{esint}
\usepackage{epstopdf}
\usepackage{braket}
\usepackage[ruled,lined]{algorithm2e}

\usepackage{hyperref}
\hypersetup{
  colorlinks   = true,    
  urlcolor     = blue,    
  linkcolor    = blue,    
  citecolor    = blue     
}

\makeatletter
\@ifundefined{textcolor}{}
{%
 \definecolor{BLACK}{gray}{0}
 \definecolor{WHITe}{gray}{1}
 \definecolor{ReD}{rgb}{1,0,0}
 \definecolor{GReeN}{rgb}{0,1,0}
 \definecolor{BLUe}{rgb}{0,0,1}
 \definecolor{CYAN}{cmyk}{1,0,0,0}
 \definecolor{MAGeNTA}{cmyk}{0,1,0,0}
  \definecolor{PURPLE}{cmyk}{0.4,0.8,0,0}
 \definecolor{YeLLOW}{cmyk}{0,0,1,0}
}

\def\d{{\rm d}}
\def\c{{\rm c}}
\def\q{{\rm q}}
\def\R{{\rm R}}
\def\ihat{{\hat{\imath}}}

\usepackage{epsfig}
\usepackage{dcolumn}
\usepackage{bm}

\makeatother

\begin{document}

\title{Bosonized one-dimensional quantum systems through enhanced event-chain Monte Carlo}

\author{Oscar Bouverot-Dupuis}
\affiliation{Universit\'{e} Paris Saclay, CNRS, LPTMS, 91405, Orsay, France}
\affiliation{IPhT, CNRS, CEA, Universit\'{e} Paris Saclay, 91191 Gif-sur-Yvette, France}
\newcommand\obd[1]{\textcolor{red}{#1}}

\author{Alberto Rosso}
\affiliation{Universit\'{e} Paris Saclay, CNRS, LPTMS, 91405, Orsay, France}

\author{Manon Michel}
\affiliation{Laboratoire de Mathématiques Blaise Pascal UMR 6620, CNRS, Université Clermont-Auvergne, Aubière, France.}
\newcommand\mm[1]{\textcolor{blue}{#1}}

\begin{abstract}
  We design an enhanced event-chain Monte Carlo algorithm to study 1D
  quantum dissipative systems, using their bosonized
  representation. Expressing the bosonized Hamiltonian as a path
  integral over a scalar field enables the application of Monte Carlo
  algorithms developed for classical systems. Specifically, we focus
  on a dissipative XXZ spin chain, exhibiting critical slowing down,
  minima degeneracy and long-range interactions. Addressing all three
  bottlenecks, we design an algorithm that combines local persistent
  event-chain Monte Carlo moves with global cluster moves, in a
  $O(1)$-complexity implementation. Through systematic performance
  analysis, we show that such an algorithm outperforms traditional
  Metropolis algorithms by more than a magnitude factor and is
  competitive with current state-of-the-art Quantum Monte Carlo
  algorithms. We then use this approach to determine the dissipative
  spin chain's phase diagram, thereby reinforcing prior analytical
  predictions.
\end{abstract}

\date{\today}

\maketitle

\section{Introduction}

The world of one dimensional (1D) quantum systems has, since nearly a century ago, intrigued by its phenomenology which sets it apart of higher dimensional quantum systems. The peculiarity of 1D quantum systems lies in the negligible role played by
particle statistics. Intuitively, this stems from the fact that
rotations used to exchange particles and probe their statistics cannot
be performed in 1D systems. This idea is at the core of the
bosonization mapping
\cite{Haldane1981LLtheory,Giamarchi,von_Delft_bosonization,1D_bosons_review,senechal_2003_Theory_methods,NdupuisCMUG2}
which provides a unifying description of bosonic and fermionic 1D
systems in terms of bosonic collective modes, hence its name. A
bosonized system can then be represented in the path integral
formalism by a classical scalar field in dimension $(1+1)$, which, for
fermions or spins, is a considerable simplification compared to the
Grassman algebra or spin coherent states usually required to treat
such problems \cite{Altland_2010_condensed}. This procedure has been
mainly used to perform renormalization group (RG) studies of the
bosonized field-theory, leading to a better understanding of various
1D quantum phenomena
\cite{Giamarchi1988Anderson1D,Schulz1986spinS,Giamarchi_1997_1DMott,Daviet_2022_flowing_bosonization}.

Among all 1D quantum systems, \emph{dissipative} quantum systems have
provided a privileged framework for studying the physics of open
quantum systems. Dissipation is typically introduced by coupling the
system to many bosonic degrees of freedom acting as a bath. The latter
are usually simple enough that they can be exactly traced out in the
path-integral formalism to yield an effective retarded interaction for
the system \cite{Caldeira_Leggett_tunneling_1981}. Dissipative quantum
dynamics were first studied in the context of systems with a single
degree of freedom such as a spin coupled to a bosonic bath in the
so-called spin-boson model
\cite{Caldeira_Leggett_spin_boson,Winter2009spinbosonMC,DeFilippis2020spinbosonMC}
or a Josephson junction coupled to a resistive environment
\cite{Schmid1983dissipation,Werner2005clusterJJ,Paris2025resilience}. More
recently, attention has been drawn to the many-body counterparts of
these dissipative systems. As previously stated, most of the
analytical progress has come from studying the bosonized Hamiltonians
\cite{Citro_2005,Lobos_2009,Cazalilla2013commensurate,bouvdup2023xxz,Long_range_dissipation}. However,
the potential of leveraging advanced computational techniques, such as
classical Monte Carlo (MC) algorithms, for a direct investigation of
the $(1+1)$D bosonized field theory remains underappreciated. Most of
the numerical effort
\cite{Weber2017retardedQMC,Weber2022diss1DQMC,ribeiro2023dissipationinduced}
put into the study of 1D dissipative quantum system has indeed been
performed directly on the microscopic Hamiltonians using standard
numerical schemes for quantum systems such as exact numerical
diagonalization \cite{Laflorencie2020BKT}, the density-matrix
renormalization group \cite{White1992DMRG,Schollowock2006DMRGreview},
or Quantum Monte Carlo (QMC) methods
\cite{Kawashima2004worldlineMC,VanHoucke2010DiagMC}. However, it is in
most cases impossible to quantitatively compare results obtained from
the microscopic and bosonized Hamiltonians as the relation between
microscopic and bosonized couplings is usually not known exactly. This
begs the question of simulating directly the bosonized field
theory. Such an approach was for instance followed to study the
resistively shunted Josephson junction \cite{Werner2005clusterJJ}
where a MC algorithm coupling global cluster moves
\cite{Sokal_1988_FK} to local Metropolis moves \cite{Metropolis1953}
was proposed to study the underlying $(0+1)$D field theory. A more
recent example can be found in the study of an XXZ spin chain subject
to local dissipation. This model, which can be seen as a many-body
generalization of the spin-boson, was studied by
\cite{bouvdup2023xxz,bouvdup2024coulombgas} with an emphasis on
analytical techniques, although a brief numerical analysis was
conducted using unadjusted Langevin dynamics. This numerical analysis
was restricted to characterizing the bulk of the phases and was not
powerful enough to probe the phase transition. The bosonized
dissipative spin chain indeed provides a prototypical example of a 1D
dissipative system hard to simulate: it
suffers from severe critical slowing-down, has an infinite number of
degenerate classical minima, and a dissipative long-range interaction
which is computationally costly to deal with.

In this work we address the simulation of the bosonized dissipative
XXZ spin chain by combining state-of-the-art classical MC algorithms
and targeting each of the three previously identified bottlenecks. We
propose an event-chain Monte Carlo (ECMC) algorithm
\cite{Michel2014GenECMC} enhanced with global cluster moves and
benefiting from a $O(1)$-complexity implementation
\cite{Michel_2019_clockMC}. The ECMC is a class of algorithms which
are non-reversible as they break the detailed balance and only satisfy
the global balance. By generating persistent and correlated moves,
ECMC algorithms produce numerical accelerations, in particular near
phase transitions, and have therefore proven useful to simulate
various particle
\cite{Bernard_2011,Michel2014GenECMC,Kapfer_2015,Isobe_2015,Klement_2021}
and spin models \cite{Michel2015spin,Nishikawa_2015}, polymers
\cite{Kampmann_2015,Harland_2017}, or more recently hard disk packings
\cite{Ghimenti_2024_ECMC}. Coupling local moves to global
cluster moves is similar to overrelaxation moves and effective in
addressing the minima degeneracy bottleneck. It has already been used
to study, for example, the $\phi^4$ theory
\cite{Brower_1989_clusterphi4} or the sine-Gordon model
\cite{Hasenbuch1994_clusterSG}, with the local moves generated by
standard Metropolis updates. Here, we show how to adapt this
enhancement by cluster moves to the local persistent ECMC
moves. Finally, we exploit the factorized structure of the acceptance
rate present in both the ECMC and cluster moves to implement them at a
reduced $O(1)$-complexity cost thanks to the clock method
\cite{Michel_2019_clockMC}. Our results show that such
an enhanced-ECMC algorithm is at least comparable to state-of-the-art
QMC approaches in the simulated system sizes and
observables. Therefore, while still being straightforward to code,
such an algorithm can be used to precisely determine the phase diagram
of the dissipative spin chain.

The paper is organized as follows: Section~\ref{sec:model} introduces
the microscopic model for the dissipative XXZ spin chain and its
mapping onto a bosonic field theory. We also give a brief summary of
known results. Section~\ref{sec:algorithm} describes the developed
algorithm based on an ECMC scheme enhanced by cluster moves. The
performance of the algorithm is then tested against other standard
algorithms and a clear scaling reduction is shown. Eventually, in
Sec.~\ref{sec:simulation}, we use our algorithm to obtain a precise
phase diagram of the dissipative XXZ spin chain. A brief
discussion of the results and concluding remarks are made in
Sec.~\ref{sec:conclusion}. Throughout the article, we work with
$\hbar=k_{\rm B}=1$.

A public version of the code is available \footnote{\url{https://plmlab.math.cnrs.fr/stoch-algo-phys/ecmc/enhanced-ecmc-bosonized}}.

\section{Model}\label{sec:model}

\subsection{Microscopic model}
We study an XXZ spin chain coupled to a set of local baths as in Refs.~\cite{bouvdup2023xxz,Sap_ohmic,Sap_subohmic}. The XXZ spin chain is a 1D spin-$\frac12$ chain of N spins, total length $L$, lattice spacing $a=\frac{L}{N}$, and whose Hamiltonian is
\begin{equation}
    \hat{H}_{\rm S}=\sum_{j=1}^N J_{\rm z} \hat{S}_j^z \hat{S}_{j+1}^z-J_{\rm xy}\left( \hat{S}_j^x \hat{S}_{j+1}^x+\hat{S}_j^y \hat{S}_{j+1}^y\right).
\end{equation}
Each site $j$ is coupled through $\hat{S}^z_j$ to a local set of quantum harmonic oscillators $\{\hat{X}_{\gamma,j}\}$ as
\begin{equation}
    \hat{H}_{\rm SB}=\sum_{j=1}^N \hat{S}_j^z \sum_\gamma \lambda_\gamma \hat{X}_{\gamma,j},
\end{equation}
and the baths are governed by the Hamiltonian
\begin{equation}
    \hat{H}_{\rm B}=\sum_{j=1}^N\sum_\gamma \frac{\hat{P}_{\gamma,j}^2}{2m_\gamma}+\frac{1}{2}m_\gamma\Omega_\gamma^2 \hat{X}_{\gamma,j}^2.
\end{equation}
The total Hamiltonian of the dissipative system is therefore $\hat{H}=\hat{H}_{\rm S}+\hat{H}_{\rm SB}+\hat{H}_{\rm B}$. We assume periodic boundary conditions for the spin chain and restrict our analysis to the zero magnetization sector.

\subsection{Effective field theory}
\begin{figure}
    \centering
    \includegraphics[width=7.5cm]{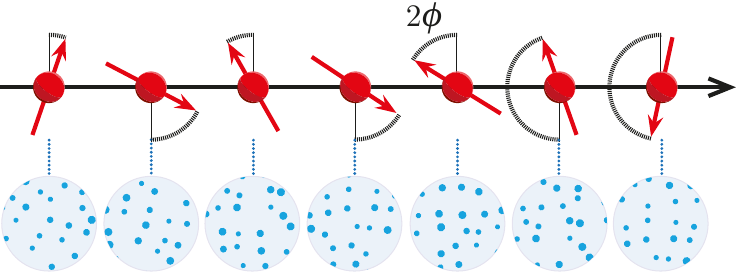}
    \caption{Schematic representation of the model. An XXZ spin chain has its spins coupled to independent and identical collections of harmonic oscillators. The orientation of the $j$-th spin is described by the angle $2\phi(x_j=ja)$ it makes with the alternating $z$-axis.}
    \label{fig:XXZ_spin_chain_drawing}
\end{figure}
This model can be rewritten as a bosonic field theory
\cite{bouvdup2023xxz} by first bosonizing the spin chain
\cite{Giamarchi,von_Delft_bosonization,NdupuisCMUG2}, and then
integrating out the bath degrees of freedom \emph{à la}
Caldeira--Leggett
\cite{Caldeira_Leggett_spin_boson,Caldeira_Leggett_tunneling_1981}. Out
of completeness, we provide in the following a more heuristic
semi-classical derivation but omit most mathematical details.

First, anticipating the fact that the dissipative XXZ spin chain has a transition to an antiferromagnetic ordered phase \footnote{This comes as no surprise since the isolated spin chain (i.e. with no bath) already exhibits a transition to an antiferromagnetic phase.}, we semi-classically represent spins through a field $\phi(x)$ as
\begin{equation}\label{eq:Sz_semiclassic}
    \hat{S}^z_j \sim (-1)^j\cos[2\phi(x_j)],
\end{equation}
with $x_j=ja$ the positions of the spins. The field $2\phi$ is simply interpreted as the angle between a spin and the alternating $z$ axis, as illustrated in Fig.~\ref{fig:XXZ_spin_chain_drawing}. Using this mapping, the canonical partition function $Z={\rm Tr}\, e^{-\beta \hat{H}}$ at inverse temperature $\beta=1/T$ can be expressed in terms of a Euclidean field theory $Z=\int \mathcal{D}\phi\, e^{-S[\phi]}$. The spin chain Hamiltonian $\hat{H}_{\rm S}$ contains interactions with coupling $J_{\rm xy}$ that want to order the spins in the $XY$ plane. Since $[\hat{S}^{x/y},\hat{S}^z]\ne 0$, this amounts to scrambling $\hat{S}^z$ (and thus $\phi$) so the associated action is
\begin{equation}
    S_{\rm LL}=\int\frac{\d x\, \d\tau}{2\pi K} \left[u\left(\partial_x\phi(x,\tau) \right)^2+\frac{1}{u}\left( \partial_\tau \phi(x,\tau)\right)^2 \right],
\end{equation}
where $x\in[0,L]$ denotes the spatial coordinate and $\tau\in[0,\beta]$ is the imaginary time coordinate. This is the so-called Luttinger liquid (LL) action with parameters $K$ and $u$. Next, the interactions in $\hat{H}_{\rm S}$ with coupling $J_{\rm z}$ are derived from Eq.~\eqref{eq:Sz_semiclassic} by noting that $\hat{S}^z_j\hat{S}^z_{j+1} \sim -\frac{1}{2}\cos[4\phi(x_j)]+{\rm cst}$. This yields
\begin{equation}
    S_g=-\frac{g}{2\pi^2}\int\frac{\d x\, \d\tau}{a\tau_c}\cos[4\phi(x,\tau)],
\end{equation}
with the short-time cutoff $\tau_c=a/u$. Finally, the baths generate an effective retarded spin-spin interaction through a dissipative kernel $K(\tau)\sim\alpha/\tau^{1+s}$ with $\alpha$ the spin-bath coupling strength and $s$ the bath exponent. Following the nomenclature introduced by Caldeira and Leggett, $s=1$ is an ohmic bath, while $0<s<1$ is a subohmic bath and $s>1$ is a superohmic bath. Since the baths are local in space and couple to $\hat{S}^z_j$, this final contribution adds up to $S_{\rm LL}+S_{\rm g}$ to give the total effective action
\begin{align}
    \label{eq:S_tot}
    S=&\int\frac{\d x\, \d\tau}{2\pi K} \left[u\left(\partial_x\phi(x,\tau) \right)^2+\frac{1}{u}\left( \partial_\tau \phi(x,\tau)\right)^2 \right]\nonumber\\
    &-\frac{g}{2\pi^2}\int\frac{\d x\, \d\tau}{a\tau_c}\cos[4\phi(x,\tau)]\\
    &-\frac{\alpha }{2\pi^2 }\underset{|\tau-\tau'|>\tau_c}{\int \frac{\d x \,\d\tau \,\d\tau'}{a\tau_c^{1-s}}}\frac{\cos[2 \phi(x,\tau)]\cos[2 \phi(x,\tau')]}{|\tau-\tau'|^{1+s}},\nonumber
\end{align}
where the last integral is over $|\tau-\tau'|>\tau_c$ to avoid any short-time divergence. This action is that of the sine-Gordon model with an additional long-range interaction that is commonly encountered in dissipative quantum systems \cite{U_Weiss_2012,Schmid1983dissipation,ribeiro2023dissipationinduced,Long_range_dissipation}. The effective parameters $u$, $K$, $g$, $\alpha$, $s$ can be related to the microscopic ones through bosonization (see \cite{bouvdup2023xxz}).

\subsection{Summary of known results}
\label{sec:known_results}
The field theory \eqref{eq:S_tot} has been previously studied by means of a renormalization group (RG) analysis \cite{bouvdup2023xxz} and a mapping to a generalized Coulomb gas \cite{bouvdup2024coulombgas}. These works showed the existence of two phases, a Luttinger liquid (LL) and an antiferromagnet (AFM), separated by a Berezinsky–-Kosterlitz–-Thouless (BKT) transition. The LL is a quasi-long-range ordered phase described by the RG fixed point
\begin{equation}\label{eq:S_LL}
    S_{\rm LL}[\phi]=\int\frac{\d x\, \d\tau}{2\pi K_\R} \left[u_\R\left(\partial_x\phi \right)^2+\frac{1}{u_\R}\left( \partial_\tau \phi\right)^2 \right],
\end{equation}
where $u_\R$, $K_\R$ are renormalized couplings and $\phi$ is implicitly evaluated at $(x,\tau)$. This phase is expected from RG arguments to exist for $K_\R\ge K_\R^c={\rm max}(1/2,1-s/2)$. For $K_\R<K_\R^c$, the system becomes an AFM and is well-approximated by the massive action
\begin{align}\label{eq:S_AFM}
    S_{\rm AFM}[\phi]=\int&\frac{\d x\, \d\tau}{2\pi K_\R}\Bigg[u_\R\left(\partial_x\phi \right)^2+\frac{\left( \partial_\tau \phi\right)^2}{u_\R}+\frac{u_\R}{\xi^2} \phi^2 \Bigg],
\end{align}
where $\xi$ is the correlation length which sets the asymptotic decay of the two-point function $\langle \phi(x,\tau)\phi(0)\rangle_{\rm AFM}\sim e^{-\sqrt{x^2+(u_\R\tau)^2}/\xi}$.

From Eq.~\eqref{eq:Sz_semiclassic}, it appears natural to define the staggered
magnetization averaged over space and imaginary-time, as the
operator
\begin{equation}\label{eq:m}
    m=\overline{\cos(2\phi(x,\tau)-2\overline{\phi(x,\tau)})},
\end{equation}
where
$\overline{f(x,\tau)}=(L\beta)^{-1}\int \d x \,\d\tau\,f(x,\tau)$. It
was shown that $m$ acts as an (infinite-order) order parameter
\cite{bouvdup2023xxz}. The substraction of the average
$\overline{\phi(x,\tau)}$ ensures the existence of
$\langle m \rangle$, in spite of the ill-defined nature of the LL
where $\phi$ is formally defined up to a constant. Using
Eqs.~(\ref{eq:S_LL},\ref{eq:S_AFM}), the finite size scaling of the
order parameter $\langle m \rangle $, where
$\langle O[\phi] \rangle=Z^{-1}\int\mathcal{D}\phi\,
O[\phi]e^{-S[\phi]}$, can be established in both phases for a system
of size $L\times\beta$ with $\beta \simeq L$,
\begin{align}\label{eq:finite_size_m_LL}
    \langle m \rangle_{\rm LL}= &\left(\frac{L_0}{L}\right)^{K_\R}\xrightarrow[L \to \infty]{}0,\\ \label{eq:finite_size_m_AFM}
    \langle m \rangle_{\rm AFM}=&\left(\frac{L_0}{\xi}\right)^{K_\R},
\end{align}
where $L_0$ is a short-distance cutoff that does not depend on $K_\R$.

\subsection{Discretized field-theory}
\label{sec:discr_FT}
For the field theory \eqref{eq:S_tot} to be well-defined, one has to introduce a short-distance cutoff \cite{Tong2018GaugeTheory} which we do by putting the field theory on a lattice. Since we want to investigate the zero-temperature and thermodynamic ($\beta,L \to \infty$) properties of the system, we consider a finite system of dimensions $L\times \beta$ as large as possible. Moreover, because the phase transition of the model belongs to the BKT universality class which has a dynamical exponent $z=1$, we take $L=u \beta$ to mitigate finite-size effects. The continuous field $\phi(x,\tau)$ is then replaced by its discretized version $\phi_i$ such that, with $\hat{x}$ and $\hat{\tau}$ the unit vectors in space and time directions,
\begin{equation}
    \phi_{i=n\hat{x}+m\hat{\tau}}=\phi(x=n a, \tau=m\tau_c),\, i\in \llbracket 1, N\rrbracket^2,
\end{equation}
with $N=L/a=\beta/\tau_c$. The periodic boundary conditions require that $\phi_{i+N \hat{x}}=\phi_i$ and $\phi_{i+N\hat{\tau}}=\phi_i$. With this discretized field, the action \eqref{eq:S_tot} becomes
\begin{align}\label{eq:S_discretized}
    S(\phi)=&\sum_i\frac{1}{2\pi K}\left[ \left(\phi_i-\phi_{i+\hat{x}} \right)^2+\left(\phi_i-\phi_{i+\hat{\tau}} \right)^2\right]\nonumber\\
    &-\sum_i\frac{g}{2\pi^2} \cos(4\phi_i)\\
    &-\sum_i\sum_{\substack{k=-\lfloor N/2 \rfloor\\k\neq 0}}^{\lfloor (N-1)/2 \rfloor} \frac{\alpha }{2\pi^2 }\frac{\cos(2 \phi_i)\cos(2 \phi_{i+k\hat{\tau}})}{k^{1+s}}\nonumber\\
    \equiv&\frac12 \sum_i \left[\sum_{j\sim i} S^{i,j}_\q(\phi) + \sum_{k=-\lfloor N/2 \rfloor}^{\lfloor (N-1)/2 \rfloor} S^{i,i+k\hat{\tau}}_\c(\phi)\right] \nonumber,
\end{align}
where  $S_\q^{i,j}(\phi)$ (resp. $S_\c^{i,j}(\phi)$) is the pairwise quadratic (resp. cosine) interaction, $j\sim i$ denotes the nearest neighbors $j$ of $i$ and the bounds of the sum over $k$ correspond to the system's periodicity.

The equilibrium probability distribution is then $\pi(\phi)=e^{-S(\phi)}/Z$ with the partition function $Z$ defined as $Z=\int \d\phi\, e^{-S(\phi)}$ with $\d\phi=\prod_i \d\phi_i$ the measure over fields. Although the partition function is formally divergent, $\phi$ being defined up to a constant, this is not an issue for our sampling procedure (see Appendix.~\ref{appendix:restricting_conf_space} for more details).

\section{Algorithm}
\label{sec:algorithm}
An efficient algorithm sampling according to the action \eqref{eq:S_discretized} must overcome three main computational bottlenecks, from both computational complexity and dynamical origins. In order to show how they arise in practice, let us consider a usual classical MCMC scheme sequentially generating single-site updates, e.g. $\phi_i \to \phi_i + \varepsilon$ with $i$ randomly picked among all sites and $\varepsilon$ some random increment (possibly depending on $\phi_i$) such that proposing $\phi_i \to \phi_i + \varepsilon$ and $\phi_i +\varepsilon \to \phi_i$ are equiprobable. At each step, such an update is accepted according to the following Metropolis acceptance ratio,
\begin{multline}\label{eq:metro-acc}
   p_{\text{Met}}(\phi, (i, \varepsilon))= \exp\Bigg(-\Bigg[\sum_{j\sim i}\Delta_\varepsilon S^{i,j}_\q(\phi)\\
   + \sum_{k=-\lfloor N/2 \rfloor}^{\lfloor (N-1)/2 \rfloor}\Delta_\varepsilon S^{i,i+k\hat{\tau}}_\c(\phi)\Bigg]_+\Bigg),
\end{multline}
with $[a]_+=\max(0,a)$ the positive part function and the $\Delta_\varepsilon$ variations of the quadratic/cosine pairwise interactions $S_{\q/\c}^{i,j}$ are
\begin{align}\label{eq:DeltaS1}
    &\Delta_\varepsilon S^{i,j}_\q(\phi)= \frac{\varepsilon^2}{2\pi K}+\frac{\varepsilon}{\pi K}\big(\phi_i-\phi_j\big),\\
    \label{eq:DeltaS2}
    &\Delta_\varepsilon  S^{i,i+k\hat{\tau}}_\c(\phi)= \delta_{k,0}\frac{g}{\pi^2}\sin(2\varepsilon)\sin(4\phi_i+2\varepsilon)\\
    &+(1-\delta_{k,0})\frac{2\alpha}{\pi^2 }\sin(\varepsilon) \frac{\sin(2 \phi_i+ \varepsilon) \cos(2 \phi_{i+k\hat{\tau}})}{|k|^{1+s}},\nonumber
\end{align}
with $\delta_{k,l}$ the Kronecker delta.
This algorithm exhibits the following bottlenecks:\vspace{0.1cm}

{\noindent (\emph{1st bottleneck})} Independent and local single-site updates most often result in diffusive dynamics that converge slowly. This issue is particularly severe in presence of strong and long-range correlations that are reinforced by the vicinity of the phase transition.\\

{\noindent (\emph{2nd bottleneck})} Multimodality, arising from the symmetry $S(\phi) = S(\phi + n\pi/2)$, $n\in \mathbbm{Z}$ further exacerbates the slow convergence of the diffusive process. \\

{\noindent (\emph{3rd bottleneck})} Computing the acceptance probability \eqref{eq:metro-acc} for the bath-induced long-range interactions $S^{i,i+k\hat{\tau}}_\c$ requires $O(N)$ operations.\\

To tackle these three bottlenecks, we propose a variant of the non-reversible event-chain Monte Carlo (ECMC) algorithm \cite{Michel2014GenECMC} enhanced by a coupling to cluster-type moves. This algorithm significantly decreases critical slowing down, although not completely suppressing it. We furthermore leverage the \emph{strict-extensive} nature of the long-range interactions to implement both types of moves in a $O(1)$ computational complexity \cite{Michel_2019_clockMC}. The key idea underlying these three elements is to consider a
\emph{factorized} Metropolis acceptance probability \cite{Michel2014GenECMC}, namely
\begin{align}\label{eq:fmetro-acc}
  p_{\text{FMet}}&(\phi, (i, \varepsilon))
  = \prod_{j\sim i} \exp\big(-[\Delta_\varepsilon S^{i,j}_\q(\phi)]_+\big) \nonumber\\
  &\times \prod_{k=-\lfloor N/2 \rfloor}^{\lfloor (N-1)/2 \rfloor}\exp\big(-[\Delta_\varepsilon S^{i,i+k\hat{\tau}}_\c(\phi)]_+\big).
\end{align}
The factorized Metropolis filter is a valid acceptance ratio, as it obeys detailed balance. Although it yields on average a smaller acceptance rate than the standard Metropolis one, its product structure makes the interactions independent. Here, this factorization allows to treat separately the different interaction terms, which is key for efficiently implementing non-reversible schemes such as ECMC \cite{Michel2014GenECMC}, cluster moves \cite{Sokal_1988_FK} and general computational complexity reduction schemes \cite{Michel_2019_clockMC}. In the following, we detail the three points and eventually compare the performance of our enhanced ECMC algorithm with a standard Metropolis algorithm, with and without enhancement by cluster moves, and a standard ECMC algorithm.

\subsection{ECMC}
\label{sec:ECMC}
\begin{figure}
    \centering
    \includegraphics[width=7.5cm]{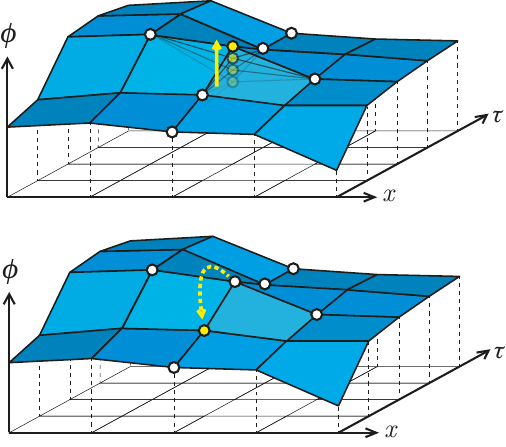}
    \caption{The two elementary steps of the ECMC algorithm. Top: a field component $\phi_i$ (in yellow) performs a ballistic shift until an event is triggered by one of its neighbors drawn in white. Bottom: when an event occurs, the neighbor involved in the event becomes the new moving field component.}
    \label{fig:ECMC_principle}
  \end{figure}
  
\subsubsection{Generating a non-reversible and rejection-free dynamics}

An ECMC algorithm \cite{Michel2014GenECMC} is a continuous-time, rejection-free and irreversible sampling scheme.  While directly identifying with a continuous-time piecewise deterministic Markov process (PDMP) \cite{Monemvassitis_2023}, such algorithms were first introduced as the infinitesimal limit $\varepsilon \to 0$ of a persistent Markov chain based on successive fixed $(i,\varepsilon)$ updates coupled to a factorized Metropolis acceptance rate (as in Eq.~\eqref{eq:fmetro-acc}).  Infinitesimal updates $(i,\varepsilon)$ are proposed until a rejection eventually happens and the update direction is resampled and given another value $(j,\varepsilon')$, so that the correct equilibrium distribution is left invariant.

Such schemes were first introduced to upgrade the dynamics from diffusive and slow ones, as typically generated by Metropolis schemes, which is exactly our first bottleneck. However, the ECMC and the linked PDMP formalism allow for great versatility, which can be exploited to couple such continuous local moves to global discrete ones, in order to address the second bottleneck linked to the multimodality. This is discussed in Section \ref{sec:cluster}.

For a general system, we recall that an ECMC algorithm is formally built from factorized $(i,\varepsilon)$ Metropolis updates following three distinct steps:
\begin{enumerate}
    \item the state space $\Omega$ is augmented to $\Omega\times\mathcal{V}$ where $v=(i,e)\in\mathcal{V}$ encodes the current ballistic motion with $e$ the sign of $\varepsilon=e|\varepsilon|$. The target distribution over the augmented space is taken of the form $\phi(\phi)\mu(v)$ in order to retrieve $\pi(\phi)$ easily. The additional variable $v$ is commonly referred to as a \emph{lifting} variable \cite{Diaconis2000lifting}.
    \item the infinitesimal update regime $\varepsilon \to 0$ limit is considered. The binomial law generating a rejection from the factorized acceptance rate $p_{\rm FMet}(\phi,(i,\varepsilon))$ transforms into a Poisson process of rate
    \begin{equation}
        \lambda(\phi, (i,e))=\lim_{\varepsilon \to 0}\frac{1-p_{\text{FMet}}(\phi, (i, \varepsilon))}{|\varepsilon|}.
    \end{equation}
    \item rejections according to the rate $\lambda(\phi,(i,e))$ trigger $v$-updates, called events, that ensure the global balance condition in the augmented state space $\Omega \times \mathcal{V}$.
\end{enumerate}
Since the introduction of the ECMC method, it has been successfully applied to different systems thanks to the flexibility enabled by the freedom of choice of the lifting variable and its update scheme. Adapting this procedure to our system, the lifting variable is $v=(i,e)\in \mathcal{V}=\llbracket 1,N\rrbracket^2 \times\{-1,1\}$ and we chose for $\mu(v)$ to be the uniform distribution over $\mathcal{V}$. In the infinitesimal regime $\varepsilon \to 0$, the factorized acceptance probability \eqref{eq:fmetro-acc} gives rise to the Poisson rate,

\begin{align}\label{eq:rate}
\lambda(\phi, (i,e))=&\lim_{\varepsilon \to 0}\frac{1-p_{\text{FMet}}(\phi, (i, \varepsilon))}{|\varepsilon|} \\
=& \sum_{j \sim i}\lambda^{i,j,e}_\q(\phi) 
  +\sum_{k=-\lfloor N/2 \rfloor}^{\lfloor (N-1)/2 \rfloor}\lambda^{i,i+k\hat{\tau},e}_\c(\phi),\nonumber
\end{align}
with $\lambda^{i,j,e}_{\q/\c}(\phi) = \lim_{\varepsilon \to 0} \frac{[\Delta_\varepsilon S_{\q/\c}^{i,j}]_+}{|\varepsilon|}=\big[e\,\partial_{\phi_i} S^{i,j}_{\q/\c}(\phi)\big]_+$ such that, from Eqs.~(\ref{eq:DeltaS1},\ref{eq:DeltaS2}),
\begin{align}\label{eq:DS}
    &\partial_{\phi_i} S^{i,j}_\q(\phi)= \frac{1}{\pi K}\big(\phi_i-\phi_j\big)\\
    &\partial_{\phi_i}  S^{i,i+k\hat{\tau}}_\c(\phi)= \delta_{k,0}\frac{2g}{\pi^2}\sin(4\phi_i)\nonumber\\ \label{eq:DS_cos}
    &\hspace{1cm}+(1-\delta_{k,0})\frac{2\alpha}{\pi^2 } \frac{\sin(2 \phi_i)\cos(2 \phi_{i+k\hat{\tau}})}{k^{1+s}},
\end{align}
Thanks to the factorization \eqref{eq:fmetro-acc}, this Poisson
process is a superposition of Poisson sub-processes of rates
$\lambda^{i,j,e}_{\q/\c}$ corresponding to the different interaction
terms. Therefore, an event triggered by the total Poisson process of
rate $\lambda$ corresponds to the event of a single interaction
term. As a rejection can be understood as an energy increment excess,
a correct redirection $v\to v'$ should ensure a balanced
redistribution of this excess. Here, and as is usually done in ECMC schemes \cite{Michel2014GenECMC,Michel2015spin}, we first leverage the pairwise symmetry
\begin{align}\label{eq:symmetry_q_term}
   [e\partial_{\phi_i} S^{i,j}_\q(\phi)]_+ = [-e\partial_{\phi_j} S^{i,j}_\q(\phi)]_+,
\end{align}
which relates the energy increments obtained by varying $\phi_i$ and $\phi_j$, so as to propose
\begin{equation}\label{eq:event}
    (i,e) \to    (j,e) \ \text{for an event triggered by } \lambda^{i,j,e}_{\q}.
\end{equation}
For the events triggered by $\lambda^{i,i+k\hat{\tau},e}_{\c}$, a
similar pairwise symmetry does not exist. However, we
  can exploit the probability flow conservation itself
  \cite{Michel2020math,Guyon_2023}. It relies here on the invariance of
  the uniform distribution for $v=(i,e)$ under the (1D) rotational
  symmetry $e \leftrightarrow -e$, and yields,
\begin{align}\label{eq:symmetry_c_term}
      &\sum_e ([e\partial_{\phi_i} S^{i,i+k\hat{\tau}}_\c(\phi)]_+ + [e\partial_{\phi_{i+k\hat{\tau}}} S^{i+k\hat{\tau},i}_\c(\phi))]_+)\nonumber\\
      =& \sum_{e} ([-e\partial_{\phi_i} S^{i,i+k\hat{\tau}}_\c(\phi)]_++[-e\partial_{\phi_{i+k\hat{\tau}}} S^{i+k\hat{\tau},i}_\c(\phi)]_+),
\end{align}
This simple relation tells us, for instance, that an energy excess
$[e\partial_{\phi_i} S^{i,i+k\hat{\tau}}_\c(\phi)]_+$ arising from
updating $\phi_i$ along $e$ can be redistributed over opposite
$(-e)$-updates of $\phi_i$ and updates of $\phi_{i+k\hat{\tau}}$ along
$e'$, according to their respective energy decrease
$[e \partial_{\phi_i} S^{i,i+k\hat{\tau}}_\c(\phi)]_+$ and
$[-e'\partial_{\phi_{i+k\hat{\tau}}}
S^{i,i+k\hat{\tau}}_\c(\phi)]_+$. The stochastic sampling of the
updated field, whether $\phi_i$ or $\phi_{i+k\hat{\tau}}$, prevents
systematic backtracking, which would otherwise occur if $\phi_i$ were
always kept fixed, necessitating repeated $e/-e$ flips that undo
previous moves. Therefore, we propose, at events triggered by $\lambda^{i,i+k\hat{\tau},e}_{\c}$,
\begin{equation}\label{eq:event2}  
   (i,e) \to  \left\{\begin{array}{l} (i,-e) \hspace{0.2cm} \text{ with probability } p_\c^{ik}(\phi)\\
    (i+k\hat{\tau},e_\c^{ik}(\phi))   \hspace{1.cm} \text{ otherwise}\\
    \end{array} \right. ,
\end{equation}
with probability
\begin{align}
    p_\c^{ik}(\phi) = \frac{|\partial_{\phi_i} S^{i,i+k\hat{\tau}}_\c(\phi)|}{|\partial_{\phi_j} S^{i,i+k\hat{\tau}}_\c(\phi)|+|\partial_{\phi_{i+k\hat{\tau}}} S^{i,i+k\hat{\tau}}_\c(\phi)|}
\end{align}
and,
\begin{align}
  e_\c^{ik}(\phi)&=-\text{sign}(\partial_{\phi_j} S^{i,i+k\hat{\tau}}_\c(\phi)),.
\end{align}
the direction which lowers the energy. From Eq.~\eqref{eq:DS_cos},
this amounts to selecting $j=i$ with probability
\begin{align}
  p_\c^{ik}(\phi) &= \frac{|\tan(2\phi_i)|}{|\tan(2\phi_i)| +|\tan(2\phi_{i+k\hat{\tau}})|},
\end{align}
and taking $j=i+k\hat{\tau}$ otherwise.

For $k=0$, the scheme (\ref{eq:event2}) comes down to
$(i,e) \to (i,-e)$, which agrees with the trivial symmetry
$[e\partial_{\phi_i} S^{i,i}_\c(\phi)]_+ = [-(-e)\partial_{\phi_i}
S^{i,i}_\c(\phi)]_+$ analogous to Eq.~\eqref{eq:symmetry_q_term}.

Such $v$-updates (\ref{eq:event},\ref{eq:event2}) satisfy the
infinitesimal limit of the global balance, through the satisfaction of
a \emph{lifted} global balance \cite{Michel2020math,Guyon_2023},
\begin{equation}
    \lambda(\phi,(j,-e'))\!=\! \sum_{v}\lambda(\phi,v)Q((\phi,v),(j,e')),
\label{eq:liftedGB}
\end{equation}
where $Q$ is the Markov kernel summarizing all the possible
$v$-updates (\ref{eq:event},\ref{eq:event2}),
\begin{multline}\label{eq:kernel}
    Q((\phi,(i,e)),(j,e')) =\sum_{i'\sim i}\frac{\lambda^{i,i',e}_\q(\phi)}{\lambda(\phi,(i,e))}\delta_{i',j}\delta_{e,e'}\\
    +\sum_{k=-\lfloor N/2 \rfloor}^{\lfloor (N-1)/2 \rfloor}\frac{\lambda^{i,i+k\hat{\tau},e}_\c(\phi)}{\lambda(\phi,(i,e))}\big(p_\c^{ik}(\phi)\delta_{i,j}\delta_{-e,e'}\hspace{1.7cm}\\
\hspace{0.8cm}    +(1-p_\c^{ik}(\phi))\delta_{i+k\hat{\tau},j}\delta_{e_\c^{ik}(\phi),e'}\big).
\end{multline}
The validity of the lifted global balance is checked using Eqs.~(\ref{eq:symmetry_q_term},\ref{eq:symmetry_c_term}).

The ECMC thus generates a deterministic dynamics determined by $v$ that is
resampled according to \eqref{eq:kernel} following
(\ref{eq:event},\ref{eq:event2}) at \emph{events} happening at a rate
\eqref{eq:rate} (see Fig.~\ref{fig:ECMC_principle}). It has now proven
more efficient and robust to directly describe the generated Markov
process by a PDMP \cite{Monemvassitis_2023,Guyon_2023}. The entire
PDMP can be characterized through its infinitesimal generator
$\mathcal{A}$ \cite{Davis_1993}. It acts on smooth test functions $f$
as
\begin{align}
    \mathcal{A}f(\phi,v)=\lim_{t \to 0^+}\mathbbm{E}_{\varphi_t}\frac{f(\varphi_t(\phi,v))-f(\phi,v)}{t},
\end{align}
where $\varphi_t$ is the time-evolution operator generated by a PDMP
over a time $t$. Owing to the definition of the PDMP, the generator of
the considered ECMC scheme is
\begin{multline}\label{eq:generator_gen_def}
  \mathcal{A}f(\phi,v)=\underbrace{\partial_{\phi_i} f(\phi,v)e}_{\text{deterministic drift}}\\
  \!+\!\underbrace{\lambda(\phi,v)}_{\text{event rate}} \sum_{v'\in\mathcal{V}}\underbrace{Q((\phi,v),v')}_{\text{jump $v\to v'$}}\left[f(\phi,v')-f(\phi,v) \right],
\end{multline}
with $\lambda$ and $Q$ respectively defined in \eqref{eq:rate} and \eqref{eq:kernel}. The derivation of the lifted global balance \cite{Michel2020math,Guyon_2023} is straightforward as the invariance of $\pi$ is equivalent to
\begin{equation}
  \sum_{v\in \mathcal{V}}\frac{1}{2N^2}\int_{\Omega} \d \phi  \mathcal{A}f(\phi,v)\pi(\phi)=0.
  \label{eq:global_balance}
\end{equation}
The lifted global balance \eqref{eq:liftedGB} is obtained after some
integration by parts and its satisfaction is checked in
Appendix~\ref{appendix:pi_invariance}. The fact that $e$ and $-e$
balance each other out in \eqref{eq:liftedGB} is the mark of the
non-reversibility of the generated process. More details on PDMP and
ECMC can be found in \cite{Monemvassitis_2023,Guyon_2023}. In
particular, we consider here a translational transport by updating a
single field $\phi_i$ along $e$, but more general flows could be used
\cite{Guyon_2023}. Also, other factorization schemes than the one in
\eqref{eq:fmetro-acc} and/or choices of $Q$ are possible. For instance, we have explored schemes which factorize the quadratic interactions in groups of 2, 3 or 4 pairwise interactions, thus associating a single event rate to each group instead of each pairwise interaction. We have also tried to separate the cosine interaction as $2\cos(2\phi_i)\cos(2\phi_j)=\cos(2\phi_i-2\phi_j)+\cos(2\phi_i+2\phi_j)$ which, factorizing both terms separately, can be simply dealt with by proposing the energy redirection $(i,e)\to (j,e)$ for the former, and $(i,e)\to (j,-e)$ for the latter. However, these schemes were always slower than the one detailed in the above.

Finally, ECMC schemes often include a \emph{refreshment} mechanism,
where the variable $v$ is resampled at some fixed time or following
some homogeneous Poisson process. Both these schemes,
  among others, do not impact the invariance condition
  \eqref{eq:liftedGB} \cite{Monemvassitis_2023}. Such refreshment
  mechanisms may prove useful to ensure ergodicity and sometimes
  even improve on the convergence
  \cite{Michel2014GenECMC,Lei_2019,Lu_2022,Eberle_2025}. In the
following, we opted for a refreshment event every fixed $T_{\rm Ref}$,
with $T_{\rm Ref}$ much larger than the typical time between events,
but much smaller than the total simulation duration.

\subsubsection{Practical implementation}

\begin{algorithm}
\caption{Enhanced ECMC}\label{alg:ECMC}
{\bf Input} $\phi, n_{\rm sample},T_{\rm Ref}, T_{\rm Sample}, T_{\rm Cluster}$ \;
${\rm Sample}=\{\}$\;
$i\gets {\rm choice}[1,\cdots,N^2]$ \tcp*{Lifting variable}
$e\gets {\rm choice}[-1,1]$\;
$t_r \gets T_{\rm Ref}$\;
$t_s \gets T_{\rm Sample}$\;
$t_{\rm Cluster} \gets T_{\rm Cluster}$\;
\While{True}{
    $j_\q \gets \underset{j\sim i}{\rm argmin}(t_\q^{i,j} \gets$ Eq.~\ref{eq:t_q_ij})\;
    $t_\q \gets \underset{j\sim i}{\rm min}(t_\q^{i,j} \gets$ Eq.~\ref{eq:t_q_ij})\;
    $t_\c,k_\c \gets$ Alg.~\ref{alg:ECMC-complexity}\;    
    $t_{\rm min}={\rm min}(t_\c,t_\q,t_{\rm Cluster},t_r)$\;
    \uIf(\tcp*[f]{Outputting a sample}){$t_s<t_{\rm min}$}{      
        $\phi_i \gets \phi_i + e \,t_s$\;
        ${\rm Sample}\gets {\rm Sample}\cup \{\phi\}$\; 
        $\phi_i \gets \phi_i + e \,(t_{\rm min}-t_s)$\;
        $t_s\gets t_s - t_{\rm min} + T_{\rm Sample}$\;
        $n_{\rm sample} \gets n_{\rm sample}-1$\;
        \uIf{$n_{\rm sample}=0$}{{\bf Break}}}
    \uElse{$t_s\gets t_s- t_{\rm min}$\;
        $\phi_i \gets \phi_i + e \,t_{\rm min}$\;}
    \uIf(\tcp*[f]{Cosine event}){$t_\c =t_{\rm min}$}{      
        $t_r\gets t_r - t_{\rm min}$\;     
        $t_{\rm Cluster}\gets t_{\rm Cluster} - t_{\rm min}$\;
        \uIf{${\rm ran}(0,1)<p_\c^{ik}(\phi)$}{
            $e,i \gets -e,i$ \;}
        \Else{ $e,i \gets e_\c^{ik_\c}(\phi),i+k_\c\hat{\tau}$ \;}}
    \uElseIf(\tcp*[f]{Quadratic event}){$t_\q =t_{\rm min}$}{
        $t_r\gets t_r - t_{\rm min}$\;    
        $t_{\rm Cluster}\gets t_{\rm Cluster} - t_{\rm min}$\;
        $e,i \gets e,j_\q$\;}
    \uElseIf(\tcp*[f]{Cluster event}){$t_{\rm Cluster} =t_{\rm min}$}{
        $t_r\gets t_r - t_{\rm min}$\;
        $t_{\rm Cluster}\gets T_{\rm Cluster}$\;
        $\phi \gets$ Alg.~\ref{alg:cluster_clock}\;}
    \uElse(\tcp*[f]{Refreshment event}){ 
        $t_r \gets T_{\rm Ref}$\;
        $t_{\rm Cluster}\gets t_{\rm Cluster} - t_{\rm min}$\;
        $i\gets {\rm choice}[1,\cdots,N^2]$\;
        $e\gets {\rm choice}[-1,1]$\;}
}
{\bf Return} ${\rm Sample}$
\end{algorithm}

To numerically simulate this PDMP, one has to first compute the event times associated with the event rates, then perform the deterministic shift until the smallest event time, and finally update the lifting variable using the Markov kernel \eqref{eq:kernel}. When moving a field component $\phi_i$, one thus has to compute an event time for each of the 4 quadratic neighbors and for each of the $N$ cosine neighbors. A quadratic rate $[e \partial_{\phi_i}S_\q^{i,j}(\phi) ]_+$ leads to an event time $t_\q^{i,j}$ which is a random variable such that
\begin{align}
    P(t_\q^{i,j} \ge t)&=\exp \left[ -\int_0^t [e \partial_{\phi_i(t')}S_\q^{i,j}(\phi(t')) ]_+ \d t'\right],
\end{align}
where $\phi_j(t)=\phi_j + \delta_{i,j} \, e \,t$ is the field evolution imposed by the lifting variable $v=(i,e)$. It can be sampled by drawing $\nu \sim {\rm ran}(0,1)$ (the uniform distribution over $[0,1]$) and solving
\begin{align}
    \nu&=\exp \left[ -\int_0^{t_\q^{i,j}} [e \partial_{\phi_i(t)}S_\q^{i,j}(\phi(t)) ]_+ \d t\right].
\end{align}
Using the definition of $\partial_{\phi_i}S_\q^{i,j}$ in Eq.~\eqref{eq:DS} leads to
\begin{equation}\label{eq:t_q_ij}
    t_\q^{i,j}=e(\phi_j-\phi_i)+\sqrt{[e(\phi_i-\phi_j)]_+^2-2\pi K\ln \nu}.
\end{equation}

Similar analytic expressions can be found for the $N$ cosine event
times $t_\c^{i,j}$. However, computing all of these $N$ event times is
unnecessarily costly and can be drastically reduced by \emph{thinning}
the cosine interactions \cite{Michel_2019_clockMC,Lewis1979thinning},
as detailed in Section~\ref{sec:complexity} and
Alg.~\ref{alg:ECMC-complexity}.

Eventually, as the generated PDMP process is continuous in time, samples are outputted every fixed time interval $T_{\rm Sample}$, corresponding here to fixed traveled distance as $e$ is of constant norm. As for the refreshment mechanism, it is possible to output samples according to exponentially distributed random times. A pseudo-code that sums up such ECMC implementation can be found in Alg.~\ref{alg:ECMC}. It already describes the enhancement by cluster moves, introduced at fixed times $T_{\rm Cluster}$, that we present in the next Section.

\subsection{Enhancing ECMC by cluster moves}
\label{sec:cluster}

\begin{figure}
    \centering
    \includegraphics[width=7.5cm]{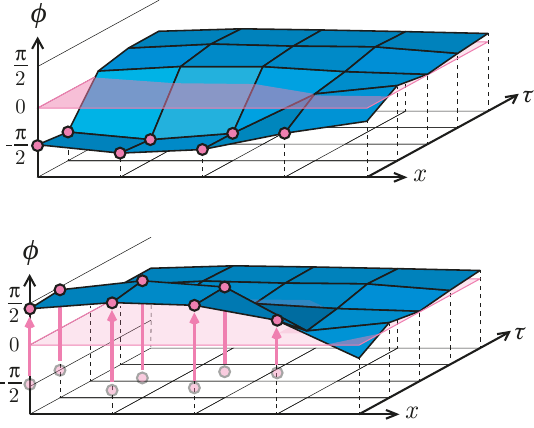}
    \caption{Basic principle of the cluster algorithm. Top: a cluster
      (pink dots) is selected and a flipping plane (in translucent
      pink) is set at one (or in between two) of the nearby potential minima located at
      $n\pi/2$, $n\in \mathbbm{Z}$. Bottom: the entire cluster is
      flipped in respect to the flipping plane.}
    \label{fig:cluster_principle}
\end{figure}

We now directly address the second dynamical bottleneck created by the symmetry $S(\phi) = S(\phi + n\pi/2)$ with $n\in\mathbbm{Z}$. It can be mitigated by efficiently moving the field from a minimum $\phi=n\pi/2$ of the action to another. The idea we propose is then to couple the continuous, rejection-free and non-reversible but local updates of standard ECMC schemes with rejection-free and global but reversible and discrete moves. Here, we leverage the symmetry $S(\phi) = S(n\pi/2- \phi)$ to design cluster moves based on the reflection
\begin{align}
    R_n\phi_i= n\pi/2-\phi_i.
\end{align}
As the $R_n$s are all involutions that leave the system globally
invariant, this enables a correct cluster algorithm
\cite{Swendsen_1987,Sokal_1988_FK}. Note that the reflection $R_n$
exploits the shift symmetry $\phi \to \phi + n\pi/2$ and the
particle-hole symmetry $\phi \to -\phi$. While the shift symmetry (at
least modulo $\pi$) is a consequence of the compact nature of the
boson $\phi$ and applies to any bosonized system, the latter
particle-hole symmetry may be absent in more general systems. This
would render the cluster algorithm inefficient, but we expect that it
could be replaced by a worm-type algorithm
\cite{Prokofiev_2001_worm}. In the following, we consider a
single-cluster construction as in the Wolff algorithm
\cite{Wolff_cluster_algo} and illustrated in
Fig.~\ref{fig:cluster_principle}.  Since the reflections $R_n$ do not
change the initial value of the field up to some modulo $\pi/2$, such
cluster moves are not ergodic on their own and are more akin to
overrelaxation moves.

We now recall the single-cluster recursive construction
\cite{Wolff_cluster_algo}. First, one uniformly picks an initial node
$i$ and adds it to the cluster. Any node $j$ interacting with
$i$ is then added to the cluster according to some probability
$p_{i,j}(\phi)$. Once every node $j$ interacting with
$i$ has been checked, the nodes interacting with the newly
added nodes are checked, and so on until the process
terminates. The entire cluster is then reflected according to
$R_n$.

The probability $p_{i,j}(\phi)$ corresponds to a rejection
of some factorized Metropolis filter. Here, we consider a
factorization different from \eqref{eq:fmetro-acc}, and group
all interactions between two fields $\phi_i,\phi_j$ together, i.e.
\begin{equation}\label{eq:P_act}
   p_{i,j}(\phi)=1-\exp(-[\Delta S^{R_n}_{\text{pair}}(\phi_i,\phi_j)]_+),
\end{equation}
where we have introduced the total pairwise action change
$\Delta S^{R_n}_{\text{pair}}(\phi_i,\phi_j)$ defined  as
\begin{multline}
  \Delta S^{R_n}_{\text{pair}}(\phi_i,\phi_j)
  = \sum_{j'\sim i}\delta_{j,j'}\Delta_\varepsilon S^{i,j'}_q(\phi)\\
  +\sum_{k=-\lfloor N/2 \rfloor}^{\lfloor (N-1)/2 \rfloor}\delta_{j,i+k\hat{\tau}}\Delta_\varepsilon S^{i,i+k\hat{\tau}}_c(\phi),
\end{multline}
with $\varepsilon = R_n\phi_i - \phi_i$.

There remains the choice of the reflection operator $R_n$, i.e. the
value $n$. A reflection operator $R_n$ is typically useful if it flips
fields $\phi_i$ such that $R_n\phi_i$ is not too far from
$\phi_i$. Indeed, if the initial field component $\phi_i$ is flipped
far away from the rest of the field, the probability $ p_{i,j}(\phi)$
will likely be large, thus adding many sites to the cluster. This will
tend to flip the entire system, resulting in a trivial, but costly,
move. We therefore condition the choice of the reflection $R_n$ on the
initial field component $\phi_i$ by uniformly picking
\begin{equation}
    n\in \llbracket2m-2,2m-1,2m,2m+1,2m+2\rrbracket,
\end{equation}
where $m=\left\lfloor \frac{2}{\pi} (\phi_i +\frac{\pi}{4})\right\rfloor$ is the
index of the minimum closest to $\phi_i$. Such a flip $R_n$ shifts
$\phi_i$ of maximum $\pm 2$ minima, which we have numerically
determined to be optimal (see Appendix~\ref{appendix:n_cluster}).  A
proof that this algorithm satisfies the detailed balance condition,
thus proving its validity, can be found in
Appendix~\ref{appendix:cluster_balance}.

Eventually, we incorporate the cluster moves into the PDMP scheme by performing them at fixed time intervals $T_{\rm Cluster}$. The PDMP then resumes by first randomly drawing the lifting variable $v$. As the cluster moves leave the correct Gibbs
measure $\pi$ invariant, such fixed-time additions to the PDMP dynamics
also leave $\pi$ invariant \cite{Monemvassitis_2023}. Furthermore, as
for the event-time generation, the long-range interactions are dealt
through the complexity-reduction clock method detailed in the next
Section. A pseudo-code that sums up the cluster generation process can
be found in Alg.~\ref{alg:cluster_clock}, along with the
complexity-reduction implementation.

\subsection{$O(1)$-complexity reduction}
\label{sec:complexity}

\begin{algorithm}
\caption{Complexity reduction for cluster generation}\label{alg:cluster_clock}
{\bf Input} $\phi$\;
$i_0 \gets {\rm choice}[1,\cdots, N^2]$ \tcp*{initial node}$C\gets \{i_0\}$ \tcp*{Add $i_0$ to cluster $C$} $S\gets \{i_0\}$  \tcp*{Add $i_0$ to the stack $S$}
$m \gets \left\lfloor \frac{2}{\pi} (\phi_{i_0} +\frac{\pi}{4})\right\rfloor$\;
$n \gets {\rm choice}[2m-2,2m-1,2m,2m+1,2m+2]$\;
\While{$S\neq \emptyset$}{
    $i\gets$ {\bf Choice}($S$)\;
    $S \gets S \setminus \{ i\}$\;
    \For(\tcp*[f]{Step 1: short-range}){$j \sim i,j \not\in C$}{
        $p_{i,j}(\phi)\gets$ Eq.~\eqref{eq:P_act}\;
        \If{${\rm ran}(0, 1) < p_{i,j}(\phi)$}{
            $C\gets C \cup \{j\}$\;
            $S\gets S \cup \{j\}$\; } }
    $n \sim f(n,\lambda^B_{\rm Clu})$ \tcp*{Step 2: long-range} \For{$l =1,2,\dots, n$}{ Pick
    $k_c\in [-\lfloor N/2 \rfloor,\lfloor (N-1)/2 \rfloor]\setminus\{-1,0,1\}$ according to $\lambda^B_{k_\c}/\lambda^B_{\rm Clu}$ using a Walker table\;
    $j\gets  i+k_\c\hat{\tau}$\;
    \If{$j \notin C$ and
        ${\rm ran}(0,1)< p_{i,j}(\phi)/p_{k_\c}^B$}{
        $C\gets C \cup \{j\}$\; $S\gets S \cup \{j\}$\; } } }
\For{$i\in C$}{$\phi_i\gets R_n\phi_i$}
{\bf Return} $\phi$\;
\end{algorithm}

\begin{algorithm}
  \caption{Complexity reduction for event generation by $\lambda^{i,j}_\c$} \label{alg:ECMC-complexity}
  {\bf Input} $(\phi, (i,e))$\;  
    $t_\c \gets 0$\;
    \While{True}{
      $\nu \sim {\rm ran}(0,1)$ \;
        $t_\c \gets t_\c -\ln \nu / \lambda^B_{\rm EC}$  \tcp*{Time to next bound event}
        Pick $k_\c \in [-\lfloor N/2 \rfloor,\lfloor (N-1)/2 \rfloor]$ according to $\lambda^B_{k_\c}/\lambda^B_{\rm EC}$ using a Walker table\;
        \If{${\rm ran}(0,1)<\lambda_\c^{i,i+k_\c\hat{\tau},e}(\phi_j + e\,t_\c\,\delta_{i,j}) /\lambda^B_{k_\c}$ }        {
            {\bf Break}\tcp*{Generated true event}
          }  }
        
 {\bf Return} $(t_\c,k_\c)$ 
\end{algorithm}

We now address the third and last bottleneck of the $O(N)$-computational complexity linked to the direct computations of the $N$ event times $t_\c^{i,j}$ in the PDMP dynamics and the $N$ interacting fields to check in the cluster moves. Both stem from the cosine interactions in the long-range dissipative term. Fortunately, both the events and cluster moves are based on factorized rates as in Eqs.~(\ref{eq:P_act},\ref{eq:rate}) which enable the use of complexity reduction schemes \cite{Lewis1979thinning,Luijten_Bloete_1995,Fukui_Todo_2009,Flores_Sola2017clusterreview} unified in the clock method \cite{Michel_2019_clockMC}. In Ref.~\cite{Michel_2019_clockMC}, it was shown that the achievable acceleration separates models into three classes defined from the strength of their long-range interactions and/or of the frustration. Our system belongs to the so-called \emph{strict-extensivity} class \cite{Michel_2019_clockMC}, i.e.
\begin{multline}
    \sum_{\substack{k=-\lfloor N/2\rfloor\\k\neq 0}}^{\lfloor (N-1)/2\rfloor} \max
    |\Delta_{\epsilon}S_c^{i,i+k\hat{\tau}}| \\\leq
    \sum_{\substack{k=-\lfloor N/2\rfloor\\k\neq 0}}^{\lfloor (N-1)/2\rfloor}
    \frac{2\alpha}{\pi^2}\frac{1}{|k|^{1+s}} \sim N^0.
\end{multline}
Therefore, we can achieve the maximal $O(N)$ to $O(1)$ (i.e. $O(N^0)$) complexity reduction. We now describe the implementation for cluster moves and event computations. Further details about the generality of the clock method and its application to a general factorized Monte Carlo procedure, including any factorized Metropolis scheme, can be found in \cite{Michel_2019_clockMC}.

\begin{figure*}[t!]
    \centering
    \includegraphics[width=1\linewidth]{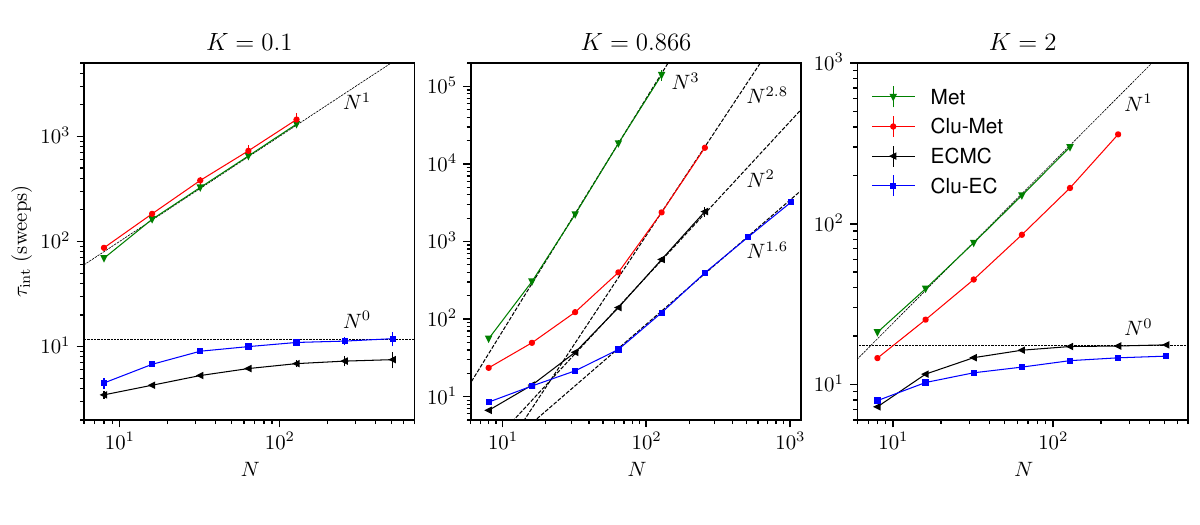}
    \caption{Integrated auto-correlation time $\tau_{\rm int}$ for the staggered magnetization $m=\overline{\cos(2\phi_i-2\overline{\phi_i})}$ deep in the AFM (left), at the phase transition (middle), and deep in the LL phase (right). The other parameters are $s=0.5$, $\alpha=1$ and $g=1$. The critical value $K=0.85$ at the transition was extracted from the results of Sec.~\ref{sec:precise_phase_diagram}. The error bars were obtained by averaging over 40 runs with independent gaussian initial conditions $\phi_i \overset{i.i.d.}{\sim} \mathcal{N}(0,1)$.}
    \label{fig:tint_scalings}
\end{figure*}

Applied to cluster moves, the clock method is used to find in $O(1)$-complexity all sites to add to the cluster from a given site $i$. From \eqref{eq:P_act}, a long-range neighbor $j$ of site $i$ is added to the cluster with probability $p_{i,j}(\phi)$. For now, $p_{i,j}(\phi)$ explicitly depends on the current field configuration $\phi$ so we need $O(N)$ operations to check all neighbors. We decide to deal with the nearest neighbors
$j \sim i$ by explicitly computing $p_{i,j}(\phi)$, and
only use a complexity-reduced implementation for distant neighbors $j=i+k\hat{\tau}$,
$|k|>1$. Following the clock method \cite{Luijten_Bloete_1995,Fukui_Todo_2009,Flores_Sola2017clusterreview,Michel_2019_clockMC}, such a distant neighbor is added to the cluster if it is first accepted with the bound probability $p^B_k \ge p_{i,i+k\hat{\tau}}$ and then accepted with the resampling probability $\frac{p_{i,i+k\hat{\tau}}}{p^B_k}$. In practice, we choose the tightest possible bound
\begin{equation}
    p_k^B = 1 - \exp\Big(-\frac{2\alpha}{\pi^2}\frac{1}{|k|^{1+s}}\Big).
\end{equation}
Since the $p^B_k$'s do not depend on $\phi$, checking the bound acceptances of all neighbors of $i$ in $O(1)$ time is now a standard procedure  \cite{Luijten_Bloete_1995,Fukui_Todo_2009,Flores_Sola2017clusterreview}. In the following, we follow the ideas of Fukui and Todo \cite{Fukui_Todo_2009} to sample the bound probabilities $p_k^B$. The neighbor $i+k\hat{\tau}$ is added with probability $p^B_k$, which is expressed as drawing an integer $n_k$ from the Poisson distribution $f_k(n_k)=\frac{e^{-\lambda_k^B} (\lambda_k^B)^{n_k} }{n_k!}$ of rate $\lambda_k^B = - \ln(1-p_k^B)=\frac{2\alpha}{\pi^2}\frac{1}{|k|^{1+s}}$ and adding site $i+k\hat{\tau}$ if $n_k >0$. Owing to the properties of Poisson distributions, the variable $n=\sum_{|k|>1} n_k$ follows a Poisson distribution of rate $\lambda^B_{\rm Clu} = \sum_{|k|>1}\lambda^B_k$. Each of the $n$ events generated is then assigned using a Walker alias table \cite{Walker1977,Marsaglia2004gen_ran_var} to the neighbor $i+k\hat{\tau}$ with probability $\lambda^B_k/\lambda^B_{\rm Clu}$. The complete procedure is thus:
\begin{itemize}
    \item Draw $n$ events from the Poisson distribution of global rate $\lambda_{\rm Clu}^B$.
    \item Assign each event to an interaction at distance $k$ ($|k|>1$) with probability $\lambda^B_k/\lambda^B_{\rm Clu}$ using a Walker table.
    \item Resample each event as actually adding site $i+k\hat{\tau}$ with probability $\frac{p_{i,i+k\hat{\tau}}}{p^B_k}$.
\end{itemize}
It is clear that the complexity identifies with the average number of events generated. This number is $\lambda_{\rm Clu}^B=\sum_{|k|>1}\frac{2\alpha}{\pi^2}\frac{1}{|k|^{1+s}} \sim O(1)$, as expected as the system belongs to the strict-extensivity class. A pseudo-code that sums up the reduction scheme for cluster generation can be found in Alg.~\ref{alg:cluster_clock}.

The clock-method is also useful for the ECMC moves to deal with the $N$ cosine events. An event generated by the rate $\lambda^{i,i+k\hat{\tau},e}_\c(\phi)$ can be interpreted as an event triggered by a bound rate $\lambda_k^B \ge \lambda^{i,i+k\hat{\tau},e}_\c(\phi)$ and then resampled with probability $\frac{\lambda^{i,i+k\hat{\tau},e}_\c(\phi)}{\lambda_k^B}$ as a true event. The tightest bound is $\lambda^B_{k\ne 0}=\frac{2\alpha}{\pi^2}\frac{1}{|k|^{1+s}}$ and $\lambda_0^B= 2g/\pi^2$. Using again the fact that the sum of Poisson processes of rates $\lambda_k^B$ is a Poisson process of rate $\lambda_{\rm EC}^B=\sum_k \lambda_k^B$, simulating this process comes down to a thinning procedure
\cite{Lewis1979thinning,Kapfer_2015,Bouchard_2018,Michel_2019_clockMC}:
\begin{itemize}
  \item Sample the first bound event time $t_\c$ following the Poisson process of rate $\lambda^B_{\rm EC}$, i.e. $t_\c = -\ln \nu /\lambda^B_{\rm EC}$ with $\nu~\sim~\text{ran}(0,1)$.
  \item Assign this bound event to an interaction at distance $k$ with probability $\frac{\lambda^B_k}{\lambda_{\rm EC}^B}$ using a Walker table.
  \item Resample this bound event as a true event with probability $\lambda^{i,i+k\hat{\tau},e}_\c(\phi) /\lambda_k^B$. If this is not a true event, go to step 1.
\end{itemize}
The complexity here identifies with the number of bound events one has to check before finding a true event. This is roughly the inverse of the average resampling probability $\sum_k \frac{\lambda_k^B}{\lambda_{\rm EC}^B}\frac{\lambda^{i,i+k\hat{\tau},e}_\c(\phi)}{\lambda_k^B}\sim O(1)$ since the bound is tight. A pseudo-code that sums up the reduction scheme for the event times can be found in Alg.~\ref{alg:ECMC-complexity}.

\begin{figure*}[t!]
    \centering
    \includegraphics[width=1\linewidth]{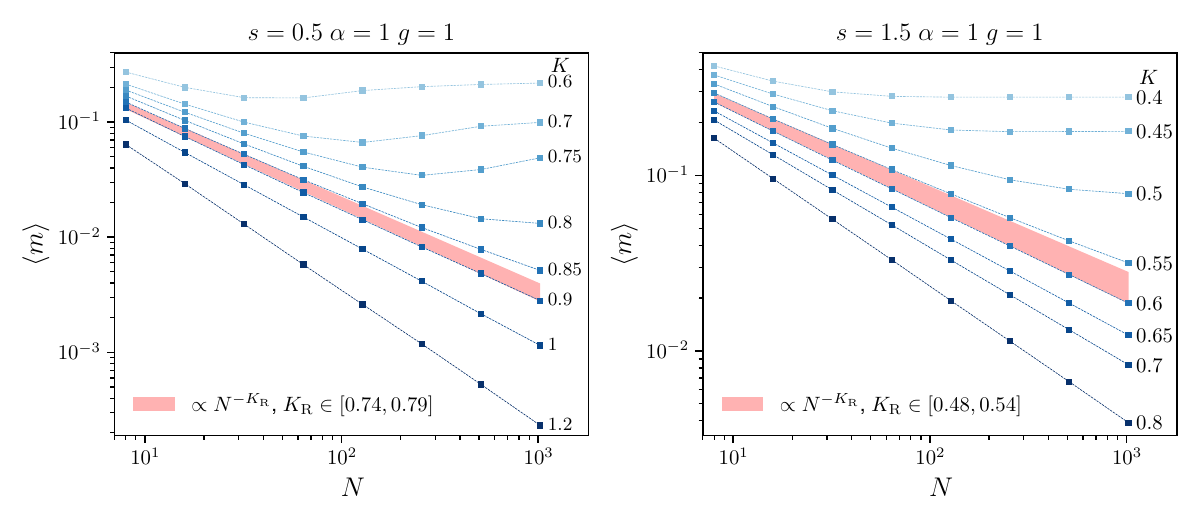}
    \caption{The averaged staggered magnetization $\langle m \rangle = \left\langle \overline{\cos(2\phi_i-2\overline{\phi_i})}\right\rangle$ as a function of the system size $N=8,16,32,64,128,256,512,1024$ for a subohmic (left) and superohmic (right) bath. The curving lines correspond to the AFM while the straight lines are in the LL. The red regions are bounded by small-$N$ fits $\propto N^{-K_\R}$ where $K_\R$ gives the renormalized Luttinger parameter. The error bars are smaller than the marker size and were obtained by averaging over 40 runs with independent gaussian initial conditions $\phi_i \overset{i.i.d.}{\sim} \mathcal{N}(0,1)$.}
    \label{fig:transition_brute_force}
\end{figure*}

\subsection{Algorithm performance}

To test the performance of our algorithm, we study the staggered
magnetization $m=\overline{\cos(2\phi_i-2\overline{\phi_i})}$
introduced in Sec.~\ref{sec:known_results}. Similarly to the situation
in spin systems \cite{Michel2015spin,Nishikawa_2015}, we expect this
observable to be a good indicator of algorithmic performance as it
quantifies all the fluctuations of $\phi$ around its average
$\overline{\phi_i}$ and should capture the slowest relaxation
dynamics, as the average itself $\overline{\phi_i}$ thermalizes
quickly thanks to the ECMC and cluster moves. We test the four
following algorithms : the Metropolis-Hastings algorithm (Met), the
Metropolis-Hastings algorithm with cluster moves (Met-Clu) as
described in Sec.~\ref{sec:cluster}, the event-chain Monte Carlo
algorithm without cluster moves (ECMC), and the event-chain Monte
Carlo algorithm with cluster moves (Clu-EC). We do not implement any complexity reduction scheme on the Met because the gain would be marginal since its moves are not persistent like in the ECMC. The algorithms including
cluster moves share their computational time equally between cluster
moves and local updates (implemented through the Met or the ECMC). In
order to draw a fair comparison between algorithms, we define the
algorithmic time $t$ expressed in sweeps (i.e. $N^2$ operations) as
increasing by 1 each time $N^2$ pairwise interactions are
evaluated. The CPU time may not be perfectly reflected in the algorithmic time but we expect both to share the same scalings and the latter to be less sensitive to specific code implementation details.

We consider the autocorrelation function defined as
\begin{align}
    C_m(t)=\frac{\langle m(t) m(0)\rangle -\langle m \rangle^2}{\langle m^2\rangle -\langle m \rangle^2}.
\end{align}
For all algorithms tested here, $C_m(t)$ decays exponentially and we
extract its characteristic time scale through the integrated
autocorrelation time
$\tau_{\rm int}=\frac{1}{2}\sum_{t=-\infty}^{+\infty}C_m(t)$. The
results for all four algorithms are shown in
Fig.~\ref{fig:tint_scalings}. Deep in both phases, the ECMC and Clu-EC
both exhibit $\tau_{\rm int}\sim N^0$ while the Met and Clu-Met have
$\tau_{\rm int}\sim N^1$. We interpret this as coming from the
complexity reduction that was implemented in the ECMC and the cluster
algorithm. Deep in the AFM ($K=0.1$), the algorithms with cluster
moves are slightly slower than those without. This comes from the fact
that the field $\phi$ gets trapped around a single minimum, so any
attempted cluster move will flip the entire field, resulting in a
trivial move. At the phase transition, the Met has
$\tau_{\rm int}\sim N^3$ and the Clu-Met seems to have a similar
large-size behavior although with a much better prefactor. This is
traced back to the diffusive dynamics of the Met which accounts for a
scaling $\sim N^2$ combined with the absence of complexity reduction
which adds an extra $\sim N^1$. Next, the integrated correlation of
the ECMC behaves as $\tau_{\rm int}\sim N^2$ which signals diffusive
dynamics with complexity reduction. Finally, the autocorrelation time
of the Clu-EC shows evidence for a $\tau_{\rm int}\sim N^{1.6}$
scaling, resulting in a clear reduction of the critical slowing down
phenomenon, in addition to complexity
reduction. It is very clear that the superior algorithm is the Clu-EC
which, at the transition, is roughly a 1000 times faster than the Met
for $N=128$, allowing for simulations up to $N=1024$ to be done in a
matter of days on a standard computer cluster.

\begin{figure*}[t]
    \centering
    \includegraphics[width=8cm]{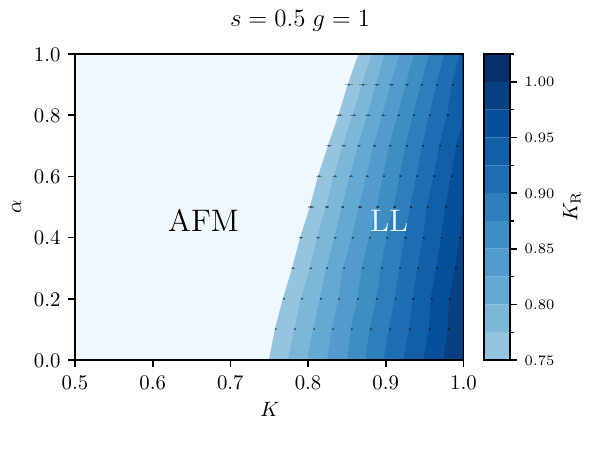}
    \includegraphics[width=8cm]{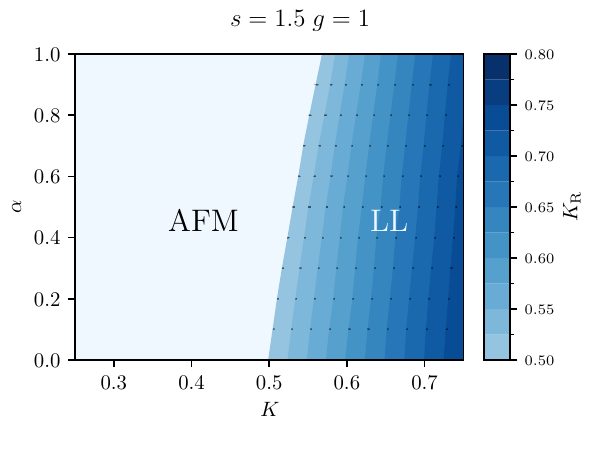}
    \caption{Left: phase diagram for a subohmic bath ($s=0.5$). Right: phase diagram for a superohmic bath ($s=1.5$). We distinguish two phases: an antiferromagnet (AFM) and a Luttinger liquid (LL). The numerical data are the dark dots. The error bars are shown but are often smaller than the marker size and were obtained using the procedure described in Fig.~\ref{fig:Kc_extraction}.}
    \label{fig:phase_diagrams}
\end{figure*}

\section{Simulation of the dissipative XXZ spin chain}
\label{sec:simulation}

We now use the ECMC algorithm with cluster moves (Clu-EC) to study the phase diagram of the dissipative XXZ spin chain introduced in Sec.~\ref{sec:model}. A first approach consists of computing the averaged staggered magnetization $\langle m \rangle$ which, using its finite-size scalings (\ref{eq:finite_size_m_LL},\ref{eq:finite_size_m_AFM}), allows to distinguish both phases. However, this approach is not the most efficient as it requires exploring large system sizes for a limited precision in return. We therefore follow a second approach where the phase transition is located from the known value of the coupling $K_\R$ at the transition, as is usually done for BKT phase transitions \cite{MCMC_XY_Weber_Minnhagen,MCMC_quantumXY_Harada,Fukui_Todo_2009}.

\subsection{Large system size determination of the phase diagram}
The averaged staggered magnetization $\langle m \rangle$ can be computed up to large sizes to determine the location of the phase transition. In Fig.~\ref{fig:transition_brute_force}, we computed $\langle m \rangle$ for a subohmic ($s=0.5$) and a superohmic ($s=1.5$) bath up to sizes $N=1024$. The data splits into two types of curves: those which in the log-log plot are lines and thus belong to the LL phase, and those which deviate from a perfect straight line and belong to the AFM. In the LL, fitting the curves gives access to the renormalized Luttinger parameter $K_\R$ from the finite size scaling $\langle m \rangle_{\rm LL}\propto N^{-K_\R}$. Fitting the data above and below the transitions gives estimates for the critical value $K_\R^c$ of $K_\R$ at the transition. For $s=0.5$, we find $K_\R^c \in [0.74,0.79]$, while for $s=1.5$ we find $K_\R^c\in [0.48,0.54]$. While the lower bounds of these estimates are fully trustworthy as they correspond to curves bending and therefore in the AFM, the upper bounds have to be considered with greater caution as they correspond to curves which remain straight up to $N=1024$ with no guarantee that this shall remain at larger $N$. Nevertheless, these results seem to confirm the RG analysis in Ref.~\cite{bouvdup2023xxz} stating that the transition lies at $K_\R^c={\rm max}(1-s/2,1/2)$, i.e. $K_\R^c=0.75$ for $s=0.5$ and $K_\R^c=0.5$ for $s=1.5$.

\subsection{Precise phase diagram}
\label{sec:precise_phase_diagram}
\begin{figure}
    \centering
    \includegraphics[trim={0.4cm 0 0.35cm 0},clip,width=7.5cm]{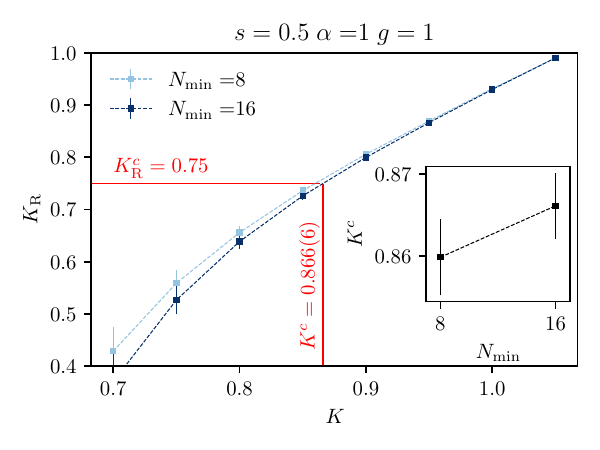}
    \caption{Extraction of $K_\R$ from the fitting of the staggered magnetization $m$ for sizes $N=8,16,32,64,128$ ($N_{\rm min}=8$) and $N=16,32,64,128$ ($N_{\rm min}=16$). The fit works well at large $K$ (LL phase) while it does not at smaller $K$ (AFM phase). The inset plot shows the value of $K^c$ obtained for both curves. Since most of the error on the determination of $K^c$ comes from the small sizes, we keep the value of $K^c$ for $N_{\rm min}=16$ and use that for $N_{\rm min}=8$ to estimate the error, leading to $K^c=0.866(6)$.}
    \label{fig:Kc_extraction}
\end{figure}
Having confirmed the validity of the RG prediction
$K_\R^c={\rm max}(1-s/2,1/2)$ in the previous section, we can now use
this knowledge to precisely determine the phase diagram. Since, in the
LL, the finite-size scaling
$\langle m \rangle_{\rm LL}=(N_0/N)^{K_\R}$ holds even for the
smallest sizes \footnote{In fact, in one of the first Monte-Carlo
studies of a BKT transition \cite{MCMC_XY_Weber_Minnhagen}, the
finite-size scaling relation was shown to be valid down to sizes
$N=3$, enabling a precise determination of the critical point from
the system sizes $N=3,4,5,6,7,8$.}, we can determine $K_\R$ only
from the system sizes up to $N=128$, saving computational time and
enabling the computations of more points on the phase diagram. For a
given set of parameters, we extract $K_\R$ by fitting the scaling of $\langle m \rangle$ for
$N=8,16,32,64,128$ and $N=16,32,64,128$ (see
Fig.~\ref{fig:Kc_extraction}). At large $K$, since the system has a well-defined LL finite-size scaling even at small sizes, the fitting of $\langle m \rangle$ works well and the determination of $K_\R$ comes with little error. At small $K$ and entering the AFM, the LL finite-size scaling does not work anymore, so extracting an apparent $K_\R$ yields large error bars. We stress that these error bars reflect the physical breakdown of the LL phase and not a poor convergence of the algorithm. For the two curves shown in Fig.~\ref{fig:Kc_extraction}, one can directly read out the critical microscopic coupling $K^c$ corresponding to
$K_\R^c={\rm max}(1-s/2,1/2)$. Since the finite-size scaling gets more accurate at large
sizes, we keep the value obtained for $N=16,32,64,128$ and compare it
with that for $N=8,16,32,64,128$ to get an estimate of the error (see
the inset in Fig.~\ref{fig:Kc_extraction}). The same procedure can be
used to extract the microscopic couplings corresponding to values of
$K_\R \ge K_\R^c$. Varying the parameters and repeating the same
analysis for each set of parameters leads to the phase diagrams shown
in Fig.~\ref{fig:phase_diagrams}. As the spin-bath coupling strength
$\alpha$ is increased, the AFM gets larger while the LL shrinks, which
is in agreement with Ref.~\cite{bouvdup2023xxz}. As a side note, the
value of the slope of the AFM-LL phase boundary at $\alpha=0$ does not
agree with the RG predictions \cite{bouvdup2023xxz}, but this was to
be expected as this slope is non-universal and depends on the exact
short-distance cutoff used, i.e. a lattice for our simulations and a
real-space hard-core repulsion between operators for the RG
computation.

\section{Discussion and conclusion}
\label{sec:conclusion}
In this work, we have proposed an enhanced-ECMC method to study
bosonized 1D quantum dissipative systems. The key idea is to work not
on the original microscopic Hamiltonian, be it fermions, bosons or
spins, but on its bosonized formulation which provides a unified
description of 1D quantum systems
\cite{Giamarchi,NdupuisCMUG2,von_Delft_bosonization}. In the path
integral formalism, a bosonized field theory is that of a simple
scalar field and thus lends itself to a study by Monte-Carlo
algorithms usually used to study classical systems. As a test model,
we have considered a dissipative XXZ spin chain analytically studied
in \cite{bouvdup2023xxz,bouvdup2024coulombgas} which is particularly
costly to simulate due to the presence of long-range interactions,
minima degeneracy and critical slowing down. This led us to design an
efficient algorithm combining local, persistent and continuous moves
from an event-chain Monte Carlo algorithm with global discrete moves
from a cluster algorithm. Both move types are relevant to address the
different dynamical bottlenecks and can be implemented in
$O(1)$-complexity. We performed a detailed performance analysis and showed
such moves result in a combined algorithm whose integrated
autocorrelation time, expressed in sweeps, scales as $O(N^{1.6})$, to
be compared to $O(N^{3})$ for a standard Metropolis algorithm.

Analytical studies of the XXZ dissipative spin chain had pointed out
the existence of two phases: a Luttinger liquid and an antiferromagnet
separated by a BKT phase transition. Thanks to the enhanced-ECMC algorithm, we were able
to first confirm the analytically predicted location of the phase
transition, and then obtain a precise picture of the global phase
diagram.

Compared to current state-of-the-art Quantum Monte Carlo (QMC)
algorithms used to study dissipative 1D quantum systems, the enhanced-ECMC algorithm
stands out as being relatively straightforward to implement while
being able to simulate system sizes up to $N=1024$, which matches or
outperforms recent QMC studies of 1D dissipative quantum systems
\cite{Weber2022diss1DQMC,Weber2017retardedQMC,ribeiro2023dissipationinduced}. It
is worth mentioning that, although we considered here a
\emph{dissipative} quantum system, this method can of course be applied
to any bosonized system. However, even if fermionic systems can be
described through bosonization, we note that such an algorithm does not
solve the sign problem in 1D as the resulting bosonic action may be
imaginary, as is the case for a spin-1 chain \cite{Schulz1986spinS}.

\vspace{1cm}

\begin{center}
{\bf Acknowledgements}
\end{center}
This work was made possible by Institut Pascal at Université
Paris-Saclay with the support of the program \emph{Investissements
d’avenir} ANR-11-IDEX-0003-01. We thank Federico Ghimenti for early
discussions about this project. We also thank Laura Foini and
Saptarshi Majumdar for fruitful discussions about the physics of
quantum dissipative systems and Alexandre Martin for great advices on
code implementation choices.  O.B.-D. acknowledges the support of the
French ANR under the grant ANR-22-CMAS-0001 (\emph{QuanTEdu-France}
project). M.M. acknowledges the support of the French ANR under the
grant ANR-20-CE46-0007 (\emph{SuSa} project). A.R. ackowledges the
support of the French ANR under the grant ANR-23-CE30-0031-04
(\emph{DISCREEP} project).

\appendix

\begin{widetext}
\section{Restricting the configuration space}
\label{appendix:restricting_conf_space}
The action $S(\phi)$ defined in Eq.~\eqref{eq:S_discretized}, the measure of fields $\d\phi$, and all physical observables (see for instance Eqs.~(\ref{eq:Sz_semiclassic},\ref{eq:m})) are invariant under the \emph{global} discrete shift symmetry $\phi \to \phi + \pi$ because the field $\phi$ is a boson compactified on a circle of radius $\pi$ \cite{Tong2018GaugeTheory}. The symmetry group $\{\phi \to \phi + n\pi, \,n\in\mathbbm{Z}\}$ having an infinite volume, the partition function is clearly divergent. What we actually want to sample are the equivalence classes $[\phi]=\{\phi'\in\mathbbm{R}^{N^2} \mid  \exists m\in \mathbbm{Z},\, \phi'=\phi+m\pi\}$ which can be represented by any of their members $\phi$ in a restricted configuration space \footnote{This is exactly like fixing a gauge in electromagnetism to get rid of the gauge symmetry redundancy.}. For instance, one can work in the space
\begin{equation}
    \Omega_\ihat = \{ \phi\in \mathbbm{R}^{N^2} \mid \phi_\ihat \in [-\pi/2,\pi/2]\}.
\end{equation}
The rest of this appendix first shows in Appendix~\ref{appendix:Z_convergence} how this ensures that the probability distribution $P(\phi)\propto e^{-S(\phi)}$ is normalizable, and then in Appendix~\ref{appendix:switching_space} why one can forget about the restricted configuration space $\Omega_\ihat$ while performing a sampling routine.

\subsection{Probability distribution normalization}
\label{appendix:Z_convergence}
This subsection aims at showing that the partition function $Z=\int_{\Omega_\ihat} \d\phi \, e^{-S(\phi)}$ is finite, making the associated probability distribution normalizable. Without loss of generality, the proof is done for the configuration space $\Omega_{0}$. We start by noticing that the action $S(\phi)$ is bounded from below by
\begin{align}
    S(\phi)&\ge\frac{1}{2\pi K}\sum_{j=1}^N\sum_{i=1}^{N-1}(\phi_{(i,j)}-\phi_{(i+1,j)})^2+\frac{1}{2\pi K}\sum_{j=1}^{N-1}(\phi_{(0,j)}-\phi_{(0,j+1)})^2-\frac{g}{2\pi^2}N^2-\frac{\alpha}{2\pi^2}N^2\sum_{j= 1}^{N-1} \mathcal{D}(j)\nonumber\\
    &=\frac{1}{2\pi K}\sum_{j=1}^N\sum_{i=1}^{N-1}n_{(i,j)}^2+\frac{1}{2\pi K}\sum_{j=1}^{N-1}m_j^2-\frac{g N^2}{2\pi^2}-\frac{\alpha N^2}{2\pi^2}\sum_{j= 1}^{N-1} \mathcal{D}(j)
\end{align}
where we have introduced $N^2-N$ space-like bond variables $n_{(i,j)}=\phi_{(i,j)}-\phi_{(i+1,j)}$ and $N-1$ time-like bond variables $m_j=\phi_{(0,j)}-\phi_{(0,j+1)}$. One can make a change of variables from the field components $\{\phi_{(i,j)}\}$ to the bond variables $\{n_{(i,j)}\}$, $\{m_j\}$ and the constrained field component $\phi_0$. Since this change of variable has a unit Jacobian,
\begin{align}
    Z&\le \int_{-\pi/2}^{\pi/2}\d \phi_0\left[\prod_{j=1}^{N}\prod_{i=1}^{N-1}\int_{-\infty}^{+\infty} \d n_{(i,j)}\,e^{-\frac{1}{2\pi K}n_{(i,j)}^2}\right] \left[\prod_{j=1}^{N}\int_{-\infty}^{+\infty} \d m_j\,e^{-\frac{1}{2\pi K}m_j^2}\right]\exp\left(\frac{g N^2}{2\pi^2}+\frac{\alpha N^2}{2\pi^2}\sum_{j= 1}^{N-1}\mathcal{D}(j)\right)\nonumber
\end{align}
which is clearly convergent.

\subsection{Switching configuration spaces while sampling}
\label{appendix:switching_space}
During the algorithm execution, the field $\phi$ might escape the configuration space $\Omega_\ihat = \{ \phi\in \mathbbm{R}^{N^2} \mid \phi_\ihat \in [-\pi/2,\pi/2]\}$. This can happen during a deterministic shift of the ECMC if the lifting variable is $(\ihat,\pm 1)$ or during a cluster flip if the site $\ihat$ gets flipped. Following Ref.~\cite{Monemvassitis_2023}, a possible fix for the ECMC part would be to add a boundary Markov kernel to trigger a global shift $\phi \to \phi \pm \pi$ whenever the boundary of $\Omega_\ihat$ is reached. For the cluster part, a similar solution can be implemented by globally shifting the field as $\phi \to \phi +n \pi$ after each cluster flip with $n\in \mathbbm{Z}$ chosen to maintain $\phi$ in $\Omega_\ihat$. However, because all observables are invariant under such a shift, we decide not to implement such eventually unnecessary boundary jumps.


\section{Global balance of the ECMC algorithm}
\label{appendix:pi_invariance}
A PDMP described by a generator $\mathcal{A}$ samples the probability distribution $\pi(\phi)$ if it is its stationary (or invariant) distribution, i.e.
\begin{align}\label{eq:pi_inv}
    \sum_{v\in \mathcal{V}}\frac{1}{2N^2}\int_{\Omega_\ihat}\mathcal{A}f(\phi,v)\pi(\phi) \,\d\phi=0.
\end{align}
In this appendix, we show, following the lines of \cite{Monemvassitis_2023}, that the probability distribution $\pi(\phi)\propto e^{-S(\phi)}$ is indeed a stationary distribution of the generator
\begin{align}\label{eq:ecmc_generator}
    \mathcal{A}f(\phi,(e,i))=&\partial_{\phi_i} f(\phi,(e,i))e\nonumber\\
    &+\sum_{j\in \partial_\q i}[e \partial_{\phi_i}S_\q^{i,j}(\phi) ]_+ \left(f(\phi,(e,j))-f(\phi,(e,i))\right)\nonumber\\
    &+\sum_{j\in \partial_\c i}[e \partial_{\phi_i}S_\c^{i,j}(\phi) ]_+\sum_{k\in\{i,j\},e'}\frac{[(-e')\partial_{\phi_k}S_\c^{i,j}(\phi) ]_+\left(f(\phi,(e',j))-f(\phi,(e,i))\right)}{\sum_{l\in\{i,j\},e''}[(-e'')\partial_{\phi_l}S_\c^{i,j}(\phi)]_+}\nonumber,
\end{align}
where all sums with $e,e',e''$ are over $\{-1,1\}$, $\partial_{\q/\c}i$ denotes the set of sites connected to $i$ through a quadratic/cosine interaction, and the remaining notations have been introduced in Sec.~\ref{sec:ECMC}. We also introduce for later use the total quadratic and cosine interactions
\begin{align}
     S_{\q/\c}(\phi)=\sum_{\langle i, j \rangle_{\q/\c}} S_{\q/\c}^{i,j}(\phi),
\end{align}
where $\langle i,j \rangle_{\q/\c}$ denotes pairs of sites connected through a quadratic/cosine interaction. Plugging the generator \eqref{eq:ecmc_generator} into Eq.~\eqref{eq:pi_inv}, one must show that the following expression vanishes
\begin{align}
   (\star) & \sum_{v\in \mathcal{V}}\frac{1}{2N^2}\int_{\Omega_\ihat}\d\phi\,\pi(\phi) \partial_{\phi_i} f(\phi,(e,i))e\nonumber\\
    (\star \star)&+\sum_{v\in \mathcal{V}}\frac{1}{2N^2}\int_{\Omega_\ihat}\d\phi\,\pi(\phi) \sum_{j\in \partial_\q i}[e \partial_{\phi_i}S_\q^{i,j}(\phi) ]_+ \left(f(\phi,(e,j))-f(\phi,(e,i))\right)\nonumber\\
    ({\star}{\star}{\star})&+\sum_{v\in \mathcal{V}}\frac{1}{2N^2}\int_{\Omega_\ihat}\d\phi\,\pi(\phi)\sum_{j\in \partial_\c i}[e \partial_{\phi_i}S_\c^{i,j}(\phi) ]_+\sum_{k\in\{i,j\}, e'}\frac{[(-e') \partial_{\phi_k}S_\c^{i,j}(\phi) ]_+\left(f(\phi,(e',k))-f(\phi,(e,i))\right)}{\sum_{l\in\{i,j\},e''}[(-e'') \partial_{\phi_l}S_\c^{i,j}(\phi) ]_+}.
\end{align}
Let us begin with the quadratic term $({\star}{\star})$. Using the pairwise symmetry $[e \partial_{\phi_i}S^{i,j}_\q]_+ = [(-e) \partial_{\phi_j} S^{i,j}_\q]_+$ leads to
\begin{align}
    ({\star}{\star})=&\sum_{e,i} \frac{1}{2N^2}\int_{\Omega_\ihat}\d\phi\,\pi(\phi) \sum_{j\in \partial_\q i}[e \partial_{\phi_i}S_\q^{i,j}(\phi) ]_+ \left(f(\phi,(e,j))-f(\phi,(e,i))\right) \nonumber\\
    =&\sum_e \frac{1}{2N^2}\int_{\Omega_\ihat}\d\phi\,\pi(\phi) \left[\sum_{j,i\in \partial_\q j}[(-e) \partial_{\phi_j}S_\q^{i,j}(\phi) ]_+ f(\phi,(e,j))- \sum_{i,j\in \partial_\q}[e \partial_{\phi_i}S_\q^{i,j}(\phi) ]_+ f(\phi,(e,i)) \right] \nonumber\\
    =&\sum_{e,i} \frac{1}{2N^2}\int_{\Omega_\ihat}\d\phi\,\pi(\phi) f(\phi,(e,i)) \sum_{j\in \partial_\q}\left[[(-e) \partial_{\phi_i}S_\q^{i,j}(\phi) ]_+ - [e \partial_{\phi_i}S_\q^{i,j}(\phi) ]_+ \right] \nonumber\\
    =&-\sum_{e,i} \frac{1}{2N^2}\int_{\Omega_\ihat}\d\phi\,\pi(\phi) f(\phi,(e,i)) \sum_{j\in \partial_\q} e \partial_{\phi_i}S_\q^{i,j}(\phi) \nonumber\\
    =&-\sum_{e,i} \frac{1}{2N^2} \int_{\Omega_\ihat}\d\phi\,\pi(\phi) f(\phi,(e,i)) e\partial_{\phi_i}S_\q(\phi),
\end{align}
where we recognized the total quadratic action $S_\q(\phi)$ in the last line. The cosine term $({\star}{\star}{\star})$ is then computed by first ordering the sums as
\begin{align}
    ({\star}{\star}{\star})=&\int_{\Omega_\ihat} \d\phi\, \pi(\phi) \frac{1}{2N^2}\sum_{e,i}\sum_{j\in \partial_\c i}\sum_{k\in\{i,j\}, e'}[e \partial_{\phi_i}S_\c^{i,j}(\phi) ]_+\frac{[(-e') \partial_{\phi_k}S_\c^{i,j}(\phi) ]_+\left(f(\phi,(e',k))-f(\phi,(e,i))\right)}{\sum_{l\in\{i,j\},e''}[(-e'') \partial_{\phi_l}S_\c^{i,j}(\phi) ]_+}.
\end{align}
Next, using the identity $\sum_i \sum_{j\in\partial_\c i}g(i)=\sum_{\langle i,j\rangle_\c}\sum_{m\in\{i,j\}}g(m)$, $({\star}{\star}{\star})$ becomes
\begin{align}
    ({\star}{\star}{\star})=&\int_{\Omega_\ihat} \d\phi\, \pi(\phi)\Bigg[\frac{1}{2N^2}\sum_{e,e'}\sum_{\langle i,j\rangle_\c}\sum_{m\in\{i,j\}}\sum_{k\in\{i,j\}}[e \partial_{\phi_m}S_\c^{i,j}(\phi) ]_+\frac{[(-e') \partial_{\phi_k}S_\c^{i,j}(\phi) ]_+f(\phi,(e',k))}{\sum_{l\in\{i,j\},e''}[(-e'') \partial_{\phi_l}S_\c^{i,j}(\phi) ]_+} \nonumber\\
    &\hspace{2cm}-\frac{1}{2N^2}\sum_{e,i}f(\phi,(e,i))\sum_{j\in \partial_\c i}[e \partial_{\phi_i}S_\c^{i,j}(\phi) ]_+\sum_{k\in\{i,j\}, e'}\frac{[(-e') \partial_{\phi_k}S_\c^{i,j}(\phi) ]_+}{\sum_{l\in\{i,j\},e''}[(-e'') \partial_{\phi_l}S_\c^{i,j}(\phi) ]_+}\Bigg]\nonumber\\
    =&\int_{\Omega_\ihat} \d\phi\, \pi(\phi)\frac{1}{2N^2}\Bigg[\sum_{e'}\sum_{\langle i,j\rangle_\c}\sum_{k\in\{i,j\}}f(\phi,(e',k)) [(-e') \partial_{\phi_k}S_\c^{i,j}(\phi) ]_+\nonumber\\
    &\hspace{2cm}-\sum_{e,i}f(\phi,(e,i))\sum_{j\in \partial_\c i}[e \partial_{\phi_i}S_\c^{i,j}(\phi) ]_+\Bigg]\nonumber\\
    =&\int_{\Omega_\ihat} \d\phi\, \pi(\phi)\frac{1}{2N^2}\Bigg[\sum_{e',k}f(\phi,(e',k))\sum_{j\in \partial_\c k} [(-e') \partial_{\phi_k}S_\c^{k,j}(\phi) ]_+ -\sum_{e,i}f(\phi,(e,i))\sum_{j\in \partial_\c i}[e \partial_{\phi_i}S_\c^{i,j}(\phi) ]_+\Bigg]\nonumber\\
    =&\int_{\Omega_\ihat} \d\phi\, \pi(\phi)\frac{1}{2N^2}\sum_{e,i}f(\phi,(e,i))\sum_{j\in \partial_\c i}\Big[ [(-e) \partial_{\phi_i}S_\c^{i,j}(\phi) ]_+ -[e \partial_{\phi_i}S_\c^{i,j}(\phi) ]_+\Big]\nonumber\\
    =&-\int_{\Omega_\ihat} \d\phi\, \pi(\phi)\frac{1}{2N^2}\sum_{e,i}f(\phi,(e,i))\sum_{j\in \partial_\c i}e \partial_{\phi_i}S_\c^{i,j}(\phi)\nonumber\\
    =&-\sum_{e,i}\frac{1}{2N^2}\int_{\Omega_\ihat} \d\phi \, \pi(\phi) f(\phi,(e,i))e \partial_{\phi_i}S_\c(\phi).
\end{align}
Putting everything together yields
\begin{align}
    \sum_{v\in\mathcal{V}}\frac{1}{2N^2}\int_{\Omega_\ihat}\mathcal{A}f(\phi,v)\pi(\phi)\,\d\phi =&\sum_{e,i}\frac{1}{2N^2}\int_{\Omega_\ihat} \d\phi \,\pi(\phi)\,e\Big[\partial_{\phi_i} f(\phi,(e,i))-f(\phi,(e,i))\partial_{\phi_i}\left(S_\q(\phi)+S_\c(\phi)\right)\Big].
\end{align}
Finally, upon using $\pi(\phi) \propto e^{-(S_\q+S_\c)}$ one arrives at
\begin{align}
    \sum_{v\in\mathcal{V}}\frac{1}{2N^2}\int_{\Omega_\ihat}\mathcal{A}f(\phi,v)\pi(\phi) \mu(v)\,\d\phi \,\d v=&\sum_{e,i}\frac{1}{2N^2}\int_{\Omega_\ihat}\d\phi \,\,e\Big[\pi(\phi)\partial_{\phi_i} f(\phi,(e,i))+f(\phi,(e,i))\partial_{\phi_i}\pi(\phi)\Big]\nonumber\\
    =&\sum_{e,i}\frac{e}{2N^2}\int_{\Omega_\ihat}\d\phi \,\partial_{\phi_i} \left[\pi(\phi)f(\phi,(e,i))\right]=0,
\end{align}
which finishes the proof.

\section{Choice of the cluster reflections}
\label{appendix:n_cluster}
When performing cluster moves as defined in Sect.~\ref{sec:cluster},
one has the freedom to choose the reflection operator $R_n$. We
argued that efficient moves are those which do not flip a field
component $\phi_i$ too far from its original position. Starting a
cluster from an initial site $i_0$, we thus propose to uniformly pick
a reflection $R_n$ among those with an index
\begin{equation}\label{eq:p_cluec}
    n\in [2m-p,2m-p+1,\cdots,2m+p-1,2m+p],
\end{equation}
where $m=\left\lfloor \frac{2}{\pi} (\phi_{i_0} +\frac{\pi}{4})\right\rfloor$ is such that $m\pi/2$ is the potential minima closest to $\phi_{i_0}$. To optimize the parameter $p$, we performed a series of runs at the phase transition for different values of $p$ and plotted the integrated autocorrelation time $\tau_{\rm int}$ of the staggered magnetization $m$ in Fig.~\ref{fig:tint_Cluecs}. The value $p=2$ seems to be the better one. Noticeably, this includes odd ($R_{2n+1}$) and even ($R_{2n}$) reflections, which was identified as a crucial performance factor for cluster algorithms applied to the sine-Gordon model \cite{Hasenbuch1994_clusterSG} or the discrete Gaussian model \cite{Evertz1991_clusterSOS}. 
\begin{figure}
    \centering
    \includegraphics[width=9cm]{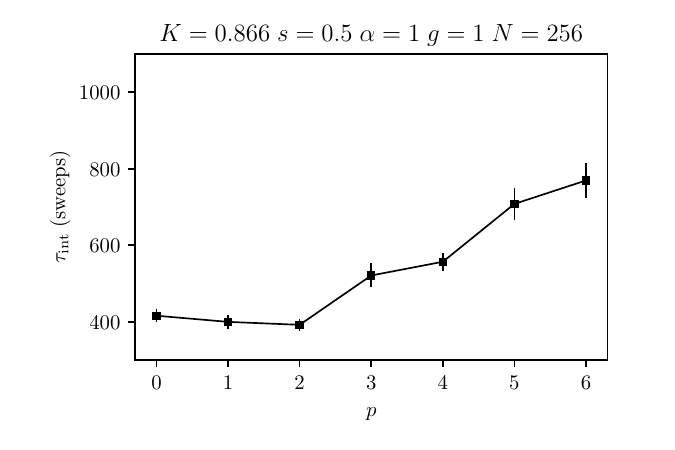}
    \caption{Integrated autocorrelation time $\tau_{\rm int}$ for the staggered magnetization $m$ with respect to the parameter $p$ defined in \eqref{eq:p_cluec} and with the Clu-EC algorithm. The error bars were obtained by averaging over 40 runs with independent gaussian initial conditions $\phi_i \overset{i.i.d.}{\sim} \mathcal{N}(0,1)$.}
    \label{fig:tint_Cluecs}
\end{figure}

\section{Detailed balance of the cluster algorithm}
\label{appendix:cluster_balance}
The cluster algorithm described in Sec.~\ref{sec:cluster} samples the correct distribution $\pi(\phi)\propto e^{-S(\phi)}$ if it satisfies the detailed balance condition between two field configurations $\phi$ and $\phi'$
\begin{align}\label{eq:detailed_balance}
    P_{\phi \to \phi'}\rm{A}_{\phi \to \phi'} \pi(\phi)=P_{\phi' \to \phi}\rm{A}_{\phi' \to \phi} \pi(\phi'),
\end{align}
where $\rm{A}_{\phi \to \phi'}$ is the probability of proposing the move $\phi \to \phi'$ and $P_{\phi\to \phi'}$ is that of accepting it. Since the cluster algorithm is rejection-free, Eq.~\eqref{eq:detailed_balance} should be satisfied with $P_{\phi\to \phi'}=P_{\phi'\to \phi}=1$. To prove this, let us first consider the probability of adding site $j$ to the cluster while looking at site $i$ with the clock implementation. For a short-range neighbor, it is by definition given by $p_{i,j}(\phi)$ (see Eq.~\eqref{eq:P_act}). For a long-range neighbor, we explicit the Fukui-Todo implementation \cite{Fukui_Todo_2009}. A long-range neighbor is added by creating $n$ bound events from $i$ with probability $f(n,\lambda^B_{\rm Clu})=e^{-\lambda^B_{\rm Clu}}{\lambda^B_{\rm Clu}}^n/n!$, attributing them to $i+k\hat{\tau}$ with probability $\lambda^B_k/\lambda^B_{\rm Clu}$, and then accepting it with probability $p_{i,i+k\hat{\tau}}(\phi)/p_k^B$ if there is at least one bound event on $i+k\hat{\tau}$. This defines the probability $p'_{i,i+k\hat{\tau}}(\phi)$ of adding a long-range neighbor as
\begin{align}
    p'_{i,i+k\hat{\tau}}(\phi)=&\sum_{n=1}^{+\infty}\frac{e^{-\lambda^B_{\rm Clu}}{\lambda^B_{\rm Clu}}^n}{n!} \sum_{n_k=1}^n \binom{n}{n_k} \left( \frac{\lambda^B_k}{\lambda^B_{\rm Clu}}\right)^{n_k}\left(1-\frac{\lambda^B_k}{\lambda^B_{\rm Clu}} \right)^{n - n_k}\frac{p_{i,i+k\hat{\tau}}(\phi)}{p_k^B}\nonumber\\
    =&\left(1-e^{-\lambda^B_{\rm Clu}} \right)\frac{p_{i,i+k\hat{\tau}}(\phi)}{p_k^B}\nonumber\\
    =& p_{i,i+k\hat{\tau}}(\phi),
\end{align}
which shows that short-range and long-range neighbors are both added with probability $p_{i,j}(\phi)$.

Keeping the previous result in mind, a cluster $C$ that relates $\phi \to \phi'$ is built by, first, selecting the correct reflection $R_n$, and then generating a tree graph $G(C)=(C,E(C))$ that originates from an initial site $i_0\in C$ and spans $C$. If one thinks of adding site $j$ to the cluster from site $i$ as activating the edge $\langle i,j\rangle$, the tree $G(C)$ is such that each edge $\langle i,j \rangle \in E(C)$ in this graph has to be accepted whereas all the edges on its border $\partial C=\{ \langle i,j\rangle \mid j\in\partial i,\, i\in C, \, j\notin C\}$ have to be rejected. Formally, this means that
\begin{align}
    A(\phi \to \phi')
    =&\sum_{i_0\in C}P(i_0 \text{ is the initial node })\sum_nP(R_n \text{ is the correct reflection}\mid i_0)\nonumber\\
    &\times\sum_{E(C)}\prod_{\langle i,j \rangle \in E(C)}p_{i,j}(\phi) \prod_{\langle i,j \rangle\in \partial C}(1-p_{i,j}(\phi)).
\end{align}
The initial node being chosen uniformly over the lattice of size $N^2$, $P(i_0 \text{ is the initial node })=1/N^2$. Furthermore, since the reflection index $n$ is conditioned on the initial node $i_0$, it can be written as $n=2m+p$ with $m=\left\lfloor \frac{2}{\pi} \phi_{i_0}+\frac{1}{2}\right\rfloor$ and $p \in \{-2,-1,0,1,2\}$. Therefore, $R_n$ is a reflection connecting $\phi$ and $\phi'$ if
\begin{align}
    n\frac{\pi}{2}-\phi_{i_0}=\phi'_{i_0} \Rightarrow p=\frac{2}{\pi}\phi_{i_0}+\frac{2}{\pi}\phi'_{i_0}-2m.
\end{align}
Writing $\phi_{i_0}=\frac{\pi}{2}(m+x)$  with $m=\left\lfloor \frac{2}{\pi} \phi_{i_0}+\frac{1}{2}\right\rfloor\in \mathbbm{Z}$ and $|x|<\frac12$ (resp. $\phi'_{i_0}=\frac{\pi}{2}(m'+x')$ with $m'=\left\lfloor \frac{2}{\pi} \phi_{i_0}+\frac{1}{2}\right\rfloor\in \mathbbm{Z}$ and $|x'|<\frac12$) leads to a reflection being valid if
\begin{align}
    p=x+x'+m'-m\Rightarrow \begin{cases}p=m'-m,\\
        x+x'=0,\end{cases}
\end{align}
which, because $p$ is uniformly chosen in $\{-2,-1,0,1,2\}$, implies
\begin{align}
    \sum_n P(R_n \text{ is the correct reflection}\mid i_0)=\frac{1}{5}P(x+x'=0)\mathbf{1}_{\{-2,\cdots,2\}}(m'-m).
\end{align}
The proposal probability $A(\phi \to \phi')$ is thus rewritten as
\begin{align}
    A(\phi \to \phi')
    =&\sum_{i_0\in C}\frac{1}{5 N^2}P(x+x'=0)\mathbf{1}_{\{-2,\cdots,2\}}(m'-m)\sum_{E(C)}\prod_{\langle i,j \rangle \in E(C)}p_{i,j}(\phi) \prod_{\langle i,j \rangle\in \partial C}(1-p_{i,j}(\phi)).
\end{align}
Putting everything together and using $p_{i,j}(\phi)=p_{i,j}(R_n \phi)$ leads to
\begin{align}
    \frac{A(\phi \to \phi')}{A(\phi' \to \phi)}
    &=\frac{\sum_{i_0\in C}P(x+x'=0)\mathbf{1}_{\{-2,\cdots,2\}}(m'-m)\sum_{E(C)}\prod_{\langle i,j \rangle \in E(C)}p_{i,j}(\phi) \prod_{\langle i,j \rangle\in \partial C}(1-p_{i,j}(\phi))}{\sum_{i_0\in C}P(x'+x=0)\mathbf{1}_{\{-2,\cdots,2\}}(m-m')\sum_{E(C)}\prod_{\langle i,j \rangle \in E(C)}p_{i,j}(\phi') \prod_{\langle i,j \rangle\in \partial C}(1-p_{i,j}(\phi'))}\nonumber\\
    &=\prod_{\langle i,j \rangle\in \partial C}\frac{1-p_{i,j}(\phi)}{1-p_{i,j}(\phi')}.
\end{align}
From the definition $p_{i,j}(\phi')=1-\exp(-[\Delta S^{R_n}_{\text{pair}}(\phi_i,\phi_j)]_+)=1-\exp(-[S(R_n \phi_i,\phi_j)-S(\phi_i,\phi_j)]_+)$ and the fact that $\forall j\notin C$, $\phi_j=\phi'_j$, one arrives at
\begin{align}
    \frac{A(\phi \to \phi')}{A(\phi' \to \phi)}
    &=\prod_{\langle i,j \rangle\in \partial C}\frac{\exp(-[\Delta S^{R_n}_{\text{pair}}(\phi_i,\phi_j)]_+)}{\exp(-[\Delta S^{R_n}_{\text{pair}}(\phi'_i,\phi'_j)]_+)}\nonumber\\
    &=\exp \left( -\sum_{\langle i,j \rangle \in \partial C} \Delta S^{R_n}_{\text{pair}}(\phi_i,\phi_j)\right)\nonumber\\
    &=\frac{\pi(\phi')}{\pi(\phi)},
\end{align}
which is the detailed balance condition \eqref{eq:detailed_balance}.

\end{widetext}

\bibliography{ref}

\end{document}